\documentclass[12pt]{article}
\usepackage[dvips]{color,graphicx}
\usepackage{hyperref}
\usepackage{amsmath,amssymb}
\usepackage{epsfig}
\usepackage{cancel}
\usepackage{appendix}
\usepackage{float}
\newcommand{\gs}{\ensuremath{g_s}} 
\newcommand{\ls}{\ensuremath{l_s}} 

\def\p{\partial}

\newcommand{\tr}{\mathop{\rm Tr}}

\def\expec#1{\langle #1 \rangle}

\newcommand{\cN}{{\mathcal{N}}}
\newcommand{\cO}{{\mathcal{O}}}

\newcommand{\trFsq}{\tr F^2}

\newcommand{\vX}{\mbox{$\vec{X}$}}

\newcommand{\Xp}{{X^{'}}}
\newcommand{\Xd}{{\dot{X}}}
\newcommand{\bsigma}{\underline{\sigma}}
\newcommand{\btau}{\underline{\tau}}

\newcommand{\be}{\begin{equation}}
\newcommand{\ee}{\end{equation}}
\newcommand{\bea}{\begin{eqnarray}}
\newcommand{\eea}{\end{eqnarray}}

\begin{document}

\begin{titlepage}

\begin{center} \Large \bf On the Beaming of  Gluonic Fields \\
 at Strong Coupling
\end{center}

\begin{center}
C\'esar A.~Ag\'on$^{\star}$\footnote{cesar.agon@nucleares.unam.mx},
Alberto G\"uijosa$^{\star}$\footnote{alberto@nucleares.unam.mx}
and Bryan O.~Larios$^{\dagger\star}$\footnote{bryanlarios@gmail.com}

\vspace{0.2cm}
$^{\star}$Departamento de F\'{\i}sica de Altas Energ\'{\i}as, Instituto de Ciencias Nucleares, \\
Universidad Nacional Aut\'onoma de M\'exico,
\\ Apartado Postal 70-543, M\'exico D.F. 04510, M\'exico\\
 \vspace{0.2cm}
$^{\dagger}\,$Departamento de F\'{\i}sica, Facultad de Ciencias, \\
Universidad Nacional Aut\'onoma de Honduras, \\
Ciudad Universitaria, Tegucigalpa, M.D.C., Honduras\\
\vspace{0.2cm}
\end{center}

\begin{center}
{\bf Abstract}
\end{center}
\noindent We examine the conditions for beaming of the gluonic field sourced by a heavy quark in strongly-coupled conformal field theories, using the AdS/CFT correspondence. Previous works have found that, contrary to naive expectations, it is possible to set up collimated beams of gluonic radiation despite the strong coupling. We show that, on the gravity side of the correspondence, this follows directly (for arbitrary quark motion, and independently of any approximations) from the fact that the string dual to the quark remains unexpectedly close to the AdS boundary whenever the quark moves ultra-relativistically. We also work out the validity conditions for a related approximation scheme that proposed to explain the beaming effect though the formation of shock waves in the bulk fields emitted by the string. We find that these conditions are fulfilled in the case of ultra-relativistic uniform circular motion that motivated the proposal, but unfortunately do not hold for much more general quark trajectories.

\vspace{0.2in}
\smallskip
\end{titlepage}
\setcounter{footnote}{0}

\tableofcontents

\section{Introduction and Summary}\label{introsec}

\subsection{Motivation}\label{motivationsubsec}

One of the most natural aspects to explore in any field theory is the manner in which field disturbances propagate. The discovery of the gauge/gravity duality \cite{malda,gkpw,magoo} has enabled this issue to be pursued for the first time in certain non-Abelian strongly-coupled gauge theories. The best understood class of examples of the correspondence relates $d$-dimensional conformal field theories (CFTs) to closed string (and therefore, gravity) theories living on a curved geometry that is asymptotically $d+1$-dimensional anti-de Sitter (AdS) times a compact space. Heavy quarks can be added to this setup by introducing open strings on the gravity side. More specifically, it is the endpoint of an open string that is dual to the quark, while its body turns out to codify the profile of the gluonic field generated by the quark. The translation between string and gauge field waves requires a computation of the bulk (gravitational, dilatonic, etc.) fields sourced by the string, focusing on their behavior near the boundary of AdS .

Even though the first analyses of this translation were carried out more than a decade ago \cite{cg,mo}, recent works have still managed to uncover a number of surprising features. In particular, whereas one might have expected radiation at strong coupling to promptly isotropize due to profuse parton branching \cite{branching}, it was found in \cite{liusynchrotron} that a quark undergoing uniform circular motion at relativistic speeds in the vacuum of the CFT (dual to the pure AdS geometry) gives rise to a gluonic field profile that essentially coincides with the familiar synchrotron spiral of classical electromagnetism. This shows that, at least for this type of trajectory, radiation is in fact \emph{beamed} along the direction of motion, and remains collimated even far away from the source.

As was emphasized in \cite{veronika}, this result is no less surprising from the gravity side of the duality, because the dominant contribution to the near-boundary bulk fields at locations that are far away from the string endpoint would be expected to originate from bits of the string that are deep inside AdS, and to be consequently spread rather than localized. Concretely, one would naively expect that after a time $\Delta t$, the information on the motion of the quark/string endpoint would have penetrated a radial coordinate distance $\sim \Delta t$ into the bulk of AdS. Nonetheless, it was shown in \cite{veronika} that, at least for the particular case of uniform circular motion, the string bits in question give rise to a localized contribution, because they move ultra-relativistically, and therefore generate bulk fields that are themselves beamed.  This observation motivated Hubeny \cite{veronika} to propose a beautiful scheme where these fields are approximated as a superposition of shock waves given off independently by each string bit. Nontrivial evidence for this proposal was provided by showing that it correctly reproduces all the qualitative and quantitative features of the exact synchrotron pattern obtained in \cite{liusynchrotron}. A particularly interesting feature of the uniform circular motion case studied in \cite{veronika} is that deep inside AdS the string bits were found to be ultra-relativistic (and thereby give rise to localized bulk fields) even when the quark is not moving at high speeds.

Hubeny's approach \cite{veronika} allows the gluonic field profile to be determined by means of a construction in terms of spatial geodesics, which is calculationally much more efficient than the traditional approach.  Significant progress could therefore be made if the same method were shown to be applicable for more general quark trajectories, as was conjectured in \cite{veronika}. If this turned out to be the case, one would ideally also like to show that the prescription is relevant in more general CFT states, such as motion of the quark through a thermal medium, which is known to be dual to a string moving outside a planar Schwarzschild-AdS black hole. Whereas the CFT vacuum is non-confining and therefore completely unlike that of QCD, the finite-temperature version of AdS/CFT has been shown to provide a useful toy model for the real-world quark-gluon plasma, in a large body of work that began with \cite{hkkky,gubser,ct,liuqhat} and has been reviewed in, e.g., \cite{clmrw}. Explorations of the gluonic field profile in this context include \cite{gluonicprofile}. The fate of a collimated beam of radiation propagating within the plasma was studied recently in \cite{rajagopalshining}.

The aim of this paper is to employ the AdS/CFT correspondence to establish how general the phenomenon of beaming is in strongly-coupled gauge theories, and in particular, to identify the conditions under  which the shock wave prescription of \cite{veronika} is valid.

\subsection{Outline and main results}\label{outlinesubsec}

Even though our analysis is expected to apply to any instance of the AdS$_{d+1}$/CFT$_{d}$ correspondence, for concreteness we phrase it in terms of the most famous and best understood example, which identifies Type IIB string theory on an asymptotically AdS$_5\times S^5$ geometry with the maximally supersymmetric (i.e., $\cN=4$) Yang-Mills theory (MSYM) on $3+1$-dimensional Minkowski spacetime.

We begin by recalling in Section \ref{mikhailovsubsec} the specifics of the quark-string connection in this context, and, in Section \ref{mappingsubsec}, the way in which the gluonic profile sourced by the quark can be determined from the shape of the dual string, with the MSYM observables $\expec{\trFsq(x)}$ and $\expec{T_{\mu\nu}(x)}$ being respectively associated with the dilatonic and gravitational fields set up by the string.

In Section \ref{originsec} we then show that, independently of any approximations, the form of the string embedding dual to the quark following an \emph{arbitrary} trajectory \cite{mikhailov} directly implies that, \emph{whenever the quark is moving ultra-relativistically, the corresponding contribution to the gluonic field will be emitted along a direction closely aligned with the velocity of the quark.} The essential reason is that this contribution is codified by a line on the string worldsheet that, as seen in (\ref{mikhbeaming}), actually remains close to the AdS boundary long after the emission event (see Fig.~1):
in a time $\Delta t$, it penetrates a radial coordinate distance $\Delta t/\gamma$ into the bulk, instead of $\Delta t$.  This therefore establishes beaming as a general feature of propagating disturbances generated by quarks in strongly-coupled gauge theories, not merely restricted to the special case of uniform circular motion studied in \cite{liusynchrotron}.

Having identified on the gravity side the physical origin of gluonic beaming for arbitrary quark trajectories, in the remainder of the paper we carry out an analysis of Hubeny's proposal \cite{veronika}. In this context, there are 2 separate issues that are of interest. First, we aim to establish whether Hubeny's approach is valid even when the quark is not ultra-relativistic, for a class of trajectories that is much more general than the synchrotron case studied in \cite{veronika}. If so, it would be useful as a calculational tool, and might imply that beaming is a generic property of gluonic radiation at strong coupling, regardless of the motion of the source. Second, in the case where the quark's motion is ultra-relativistic, the formation of shock waves as proposed by Hubeny would reinforce the beaming effect due to the string's unusual proximity to the boundary (as described in Section \ref{originsec}).

After defining and computing in Section \ref{transversesec} the transverse velocity $V_{\perp}$ of the points along the string, we work out in Section \ref{pulverizingsec} the conditions of validity for the approximation scheme proposed in \cite{veronika}. Specifically, Section \ref{fsqsubsec} identifies (\ref{xdotcondition}) and (\ref{xprimecondition}) as necessary and sufficient requirements for the string to be well approximated, as far as the dilaton field it generates and the associated $\expec{\trFsq(x)}$ are concerned, as a collection of point particles of equal mass, moving with velocities $V_{\perp}$. Section \ref{tmunusubsec} extends the analysis to the case of the graviton field and the associated $\expec{T_{\mu\nu}(x)}$, showing that a third condition is needed for the string to be decomposable into particles, equation (\ref{relativisticcondition}), which demands that the string bits move ultra-relativistically. This is the condition that was the focus of \cite{veronika}, and naturally brings into play the localization of the bulk fields along shock waves, as advocated in that work. Section \ref{shocksubsec} recalls simple facts about the way in which such shock waves are formed, highlighting the fact that they can take a substantial amount of time to build up. This leads to a fourth condition for the general applicability of the mechanism of \cite{veronika}.

The next 3 sections examine to what extent the 3 conditions needed to `pulverize' the string are satisfied for arbitrary quark trajectories, in various possible parametrizations of the string worldsheet. (The analysis of the fourth, shock wave formation, condition is deferred until the end of the paper.) A natural place to start is the  parametrization employed in the construction of the relevant string embeddings \cite{mikhailov}, which efficiently tracks the flow of quark/endpoint data along the worldsheet and is instrumental in elucidating the origin of the beaming effect on the gravity side of the correspondence. In Section \ref{mikhailovsec} we find that with this gauge choice, the 3 `pulverizability' conditions hold only for a very restricted class of quark trajectories, where the magnitude $a$ of the 4-acceleration is approximately constant. About the only memorable member of this class is precisely the case of uniform circular motion, for which $a$ is of course constant.

In Section \ref{particlesec} we try a different approach: in search of a worldsheet parametrization that might be ideally suited to the scheme of \cite{veronika}, we fix the gauge by requiring 2 of the 3 conditions to be automatically satisfied. Unfortunately, through this route, we find that (at least under the simplifying assumptions that allow us to explicitly solve for the desired worldsheet coordinates) the remaining condition fails to hold even in the case of constant $a$ that had already been satisfactorily covered in the preceding section.

We then move on in Section \ref{staticsec} to the static gauge associated with the AdS Poincar\'e coordinates (\ref{metric}), which is a natural possibility and was in fact the parametrization employed in \cite{veronika}. In Section \ref{staticsubsec} we find that if the quark is ultra-relativistic, the string generically satisfies conditions (\ref{xdotcondition}) and (\ref{relativisticcondition}) and can therefore be decomposed into particles, but in general it does not satisfy (\ref{xprimecondition}), meaning that these particles cannot be assigned equal mass. For a generic quark trajectory, the resulting particles have masses that are time-dependent, and hence do not give rise to shock waves as proposed in \cite{veronika}.  An exception occurs when the quark  moves with constant velocity, which gives a simple, time-independent scaling of the masses. Another special case is uniform circular motion, where condition (\ref{xprimecondition}) is actually satisfied, but the more important condition (\ref{xdotcondition}) is violated. Nonetheless, we show in Section \ref{semistaticsubsec} that for this special type of motion, a particular reparametrization of just the spatial worldsheet coordinate (under which $V_{\perp}$ is left unaffected) allows all 3 of the pulverizability conditions to be satisfied, which is consistent with the success of the calculation in \cite{veronika}. In Section \ref{shockformationsubsec} we finally return to the shock wave formation condition identified in Section \ref{shocksubsec}, and find that even for uniform circular motion this fourth condition is only approximately satisfied to the extent that the quark itself is ultra-relativistic.

Altogether, then, the conclusion from Sections \ref{mikhailovsec}-\ref{staticsec} is unfortunately negative: even though we are able to verify that the approximation scheme of \cite{veronika} is  justified for the case of ultra-relativistic uniform circular motion, the same chain of analysis reveals that, contrary to our hopes, the method is not applicable for a much wider class of quark trajectories. One obstacle is that, in a random  parametrization, the string cannot really be decomposed into particles moving with velocities $V_{\perp}$. Even if this hurdle is avoided by choosing an appropriate  gauge, the string bits turn out to be ultra-relativistic only in a portion of the worldsheet that is rather narrow (especially when one restricts attention to the swath of the worldsheet that can causally influence a given observation point on the boundary), unless the quark itself is ultra-relativistic. But even then, we face the problem that the masses of the particles into which the string is decomposed are generally time-dependent.  Another important obstacle is that, for arbitrary quark trajectory, the motion of each string bit is not well approximated by a null geodesic for a time long enough for the corresponding shock wave to be able to extend all the way to the AdS boundary. About the only memorable cases for which Hubeny's prescription eludes all of these difficulties are those of (ultra-relativistic) constant velocity and uniform circular motion. For these cases, the beaming effect identified in Section \ref{originsec} of the present paper, due to the proximity of the string to the AdS boundary, is enhanced by the localization of the bulk fields along shock waves, as advocated in \cite{veronika}.

\section{Preliminaries}\label{preliminariessec}

\subsection{The quark as a string}\label{mikhailovsubsec}
A heavy quark propagating in the vacuum of $SU(N_c)$ MSYM on $(3+1)$-dimensional Minkowski space is described in dual language by a string moving on the Poincar\'e wedge of the AdS$_5$ geometry,
\begin{equation}\label{metric}
ds^2=G_{mn}dx^m dx^n={R^2\over z^2}\left(
-dt^2+d\vec{x}^{\,2}+dz^2 \right)~.
\end{equation}
The coordinates $x^{\mu}\equiv(t,\vec{x})$  parallel to the AdS boundary $z=0$  are
 identified with the gauge theory spacetime coordinates, whereas the
radial direction $z$ is mapped to a variable length (or, equivalently, inverse energy) scale in MSYM
\cite{uvir}.  The MSYM coupling is connected to the string coupling via $g_{YM}^2=4\pi\gs$. The radius of curvature $R$ is related to the gauge theory 't Hooft coupling $\lambda\equiv g_{YM}^2 N_c$ through
\begin{equation}\label{lambda}
 \lambda={R^4\over \ls^4}~,\nonumber
\end{equation}
 where $\ls$ denotes the string length. Throughout this paper we will consider for simplicity the case of an infinitely massive quark, which corresponds to a string extending all the way from the Poincar\'e horizon at $z\to\infty$ to the AdS boundary at $z=0$.\footnote{The full setup includes elements that will be ignored here because they do not play any role in our analysis: a 5-sphere of radius $R$ at each point of AdS$_5$, $N_c$ units of flux through this 5-sphere of a background (Ramond-Ramond) self-dual 5-form field strength, and $N_f$ flavor D7-branes at the AdS boundary (or, more generally, extending from the boundary down to $z=\sqrt{\lambda}/2\pi m$, where $m$ is the quark mass).}

In the large $N_c$, large $\lambda$ limit, the string embedding is determined by extremizing the Nambu-Goto action
\begin{equation}\label{nambugoto}
S_{\mbox{\scriptsize NG}}=-{1\over 2\pi\ls^2}\int
d^2\sigma\,\sqrt{-\det{g_{ab}}}
=-{1\over 2\pi\ls^2}\int
d^2\sigma\,\sqrt{\left(\Xd\cdot\Xp\right)^2-\Xd^2\Xp^2}~,
\end{equation}
where $g_{ab}\equiv\p_a X^m\p_b X^n G_{mn}(X)$ ($a,b=0,1$) is
the induced metric on the worldsheet, and of course $\,\dot{}\equiv\p_{\sigma^{0}}\equiv\p_{\tau}$, ${}^{\prime}\equiv\p_{\sigma^1}\equiv\p_{\sigma}$. One of the parametrizations we will employ is the static gauge
$\tau=t$, $\sigma=z$, where the string embedding is described as $\vX(t,z)$.

The quark's trajectory coincides with the path of the string endpoint at the AdS boundary,
\begin{equation}\label{quarkx}
x^{\mu}(\tau)=X^{\mu}(\tau,\sigma)|_{z=0}~.
\end{equation}
For any timelike quark trajectory $x^{\mu}(\tau_r)$, parametrized by \emph{proper} time $\tau_r$, the dual string embedding that corresponds to a purely outgoing gluonic field configuration was found by Mikhailov \cite{mikhailov},
\begin{equation}\label{mikhsol}
 X^{\mu}(\tau_r,z)=x^{\mu}(\tau_r)+zv^{\mu}(\tau_r)~,
\end{equation}
where $v^{\mu}\equiv dx^{\mu}/d\tau_r$ denotes the 4-velocity of the quark (such that $\eta_{\mu\nu}v^{\mu}v^{\nu}=-1$). In non-covariant notation, (\ref{mikhsol}) stipulates that
\begin{eqnarray}\label{mikhsolnoncovariant}
\vec{X}(t_r,z)&=&\vec{x}(t_r)+\frac{\vec{v}(t_r) z}{\sqrt{1-\vec{v}(t_r)^2}}~,\\
t(t_r,z)&=&t_r+\frac{z}{\sqrt{1-\vec{v}(t_r)^2}}~, \nonumber
\end{eqnarray}
where we have used $d\tau_r=dt_r/\gamma$, $v^{\mu}=\gamma(1,\vec{v})$. {}From (\ref{mikhsolnoncovariant}) we see that the behavior of the string at a given time $t=X^0$ and radial depth $z$ (which essentially encodes the gluonic field a distance $z$ away from the quark) is parametrized in terms of the behavior of the quark/string endpoint at the earlier, \emph{retarded} time $t_r$, in complete analogy with the Lienard-Wiechert story in classical electrodynamics.\footnote{The way in which the preceding expressions are modified for finite quark mass has been worked out in \cite{dragtime,lorentzdirac,damping} (and reviewed in \cite{jphysg}), and it would be straightforward (if a bit tedious) to extend to that case the analysis of the present paper.}

\subsection{Mapping out the gluonic field}\label{mappingsubsec}

The gluonic field sourced by the quark can be mapped out by computing the expectation value of local MSYM operators. The simplest is the Lagrangian density operator
\begin{equation}\label{o}
\cO_{F^{2}}(x^{\mu})\equiv{1\over 2 g_{YM}^{2}}
\tr \left\{
   F^{2}(x^{\mu})+[\Phi_{I},\Phi_{J}] [\Phi^{I},\Phi^{J}](x^{\mu})
  +\mbox{fermions}
\right\}~,
\end{equation}
which we will abbreviate simply as $\expec{\trFsq(x^{\mu})}$. This has been done in \cite{dkk,cg,otherdilaton} for special cases, and in \cite{trfsq} for arbitrary quark trajectories. The operator (\ref{o}) is known to be
dual to the (s-wave) dilaton field
$\phi(x)$ in AdS \cite{igor,iwm}. The GKPW recipe for correlation functions \cite{gkpw} at large $N_c$ and $\lambda$ relates its
one-point function to a variation of the supergravity action with respect to the boundary value of $\phi$. The connection can be summarized as \cite{cg}
\begin{equation} \label{trfsq}
\expec{\trFsq(x^{\mu})}=-
\lim_{z\to 0}\left({1\over z^{3}}\p_{z}\varphi(x^{\mu},z) \right)~,
\end{equation}
in terms of a rescaled dilaton field $\varphi\equiv R^{3}\phi/ 16\pi G^{(5)}_{\mbox{\scriptsize N}}$. The latter is in turn obtained  by convolving the string source\footnote{In the absence of the string, the dilaton field is constant, so there is no real difference between the Einstein and string frame metrics, but to derive the source term in the dilaton equation of motion, one must differentiate the Nambu-Goto action with respect to the dilaton, holding the Einstein frame metric $G^{E}_{MN}\equiv e^{-\phi/2}G_{MN}$ fixed.}
\begin{equation} \label{stringsource}
J(x)\equiv {{R^3}\over (16\pi G^{(5)}_{\mbox{\scriptsize N}})^2}\frac{\delta S_{\mbox{\scriptsize NG}}}{\delta \varphi(x)}=
{1\over 4\pi\ls^2}
     \int d^2\sigma\sqrt{\left(\Xd\cdot\Xp\right)^2-\Xd^2\Xp^2}\,\delta^{(5)}\left(x-X(\tau,\sigma)\right)
\end{equation}
with the retarded dilaton propagator $D(x;\bar{x})$ \cite{dkk}:
\begin{equation} \label{dilstring}
\varphi(x)={1\over 4\pi\ls^2}\int d\tau\,d\sigma\,\sqrt{\left(\Xd\cdot\Xp\right)^2-\Xd^2\Xp^2}
  \,D(x;X(\tau,\sigma))~.
\end{equation}

A second (and more informative) way to map out the gluonic field set up by the quark is by computing the expectation value of the MSYM energy-momentum tensor, which is related to the graviton field generated by the string, $h_{mn}(x)\equiv G_{mn}^{\mbox{\scriptsize string}}(x)-G_{mn}(x)$. The calculation of $\expec{T_{00}}$ has been carried out in \cite{mo,gluonicprofile,shuryakdipole,liusynchrotron,iancu1} for special cases and in \cite{iancu2} for arbitrary quark trajectories. In a gauge where $h_{mz}=0$, and in terms of the coefficients of the near-boundary expansion
\begin{equation}\label{metricexpansion}
h_{\mu\nu}(x^{\rho},z)=\frac{R^2}{z^2}\left(
z^4 h^{(4)}_{\mu\nu}(x^{\rho})
+\ldots\right)~,
\end{equation}
the recipe can be written in the form \cite{bk,dhss}
\begin{equation}\label{graltmunu}
\expec{T_{\mu\nu}(x^{\rho})}={R^{3} \over 4\pi G^{(5)}_{\mbox{\scriptsize N}}}
h^{(4)}_{\mu\nu}(x^{\rho})~.
\end{equation}
Again, the bulk field is determined by Green's function methods,
\begin{equation} \label{hsol}
h_{mn}(x)=\int d^{5}\bar{x}\sqrt{-G}\, \mathcal{G}_{mn;\bar{m}\bar{n}}(x,\bar{x})
  \mathrm{T}^{\bar{m}\bar{n}}(\bar{x}),
\end{equation}
with $\mathcal{G}_{mn;\bar{m}\bar{n}}(x,\bar{x})$ the graviton propagator (see, e.g., \cite{dhoker}), and $\mathrm{T}^{mn}(x)$ the energy-momentum tensor of the string in AdS,
\begin{eqnarray} \label{stringtmn}
\mathrm{T}^{mn}(x)&\equiv & \frac{2}{\sqrt{-G}}\frac{\delta S_{\mbox{\scriptsize NG}}}{\delta G_{mn}(x)}
\\
{}&=&\frac{1}{2\pi\ls^2}
     \int  \frac{d^2\sigma}{\sqrt{-G}\sqrt{-g}}
  \epsilon^{ab}\epsilon^{cd}\p_{a}X^{m}\p_{c}X^{n}\p_{b}X^{p}\p_{d}X^{q}
  G_{pq}\,\delta^{(5)}\left(x-X(\tau,\sigma)\right)~.\nonumber
\end{eqnarray}

\section{The Origin of Gluonic Beaming}\label{originsec}
As recalled in section \ref{mikhailovsubsec}, Mikhailov \cite{mikhailov} was able to solve the full (non-linearized) Nambu-Goto equations to construct the string embedding (\ref{mikhsol}) (or, in non-covariant notation, (\ref{mikhsolnoncovariant})) dual to a quark with arbitrary trajectory. The resulting worldsheet is a ruled surface, spanned by lines at  constant retarded proper time $\tau_r$ (or retarded time $t_r$). Combining (\ref{metric}) and (\ref{mikhsol}), the induced metric on the worldsheet is found to be
\begin{eqnarray}\label{mikhwsmetric}
g_{\tau_r\tau_r}&=&\left(\frac{\p X}{\p \tau_r}\right)_z^2={R^2\over z^2}(a^2 z^2-1)~,
\nonumber\\
g_{zz}&=&\left(\frac{\p X}{\p z}\right)_{\tau_r}^2=0~,
\qquad \\
g_{\tau_r z}&=&\left(\frac{\p X}{\p \tau_r}\right)_z\!\!\cdot\left(\frac{\p X}{\p z}\right)_{\tau_r}
=-{R^2\over z^2}~,\nonumber
\end{eqnarray}
where $a^2\equiv a^{\mu}(\tau_r)a_{\mu}(\tau_r)$ is the norm of the 4-acceleration. {}From $g_{\tau_r\tau_r}$ we see that
\begin{equation}\label{slc}
z_*(\tau_r)\equiv 1/\sqrt{a^2}
\end{equation}
marks the location of a stationary limit curve, beyond which it is $z$ and not $\tau_r$ that plays the role of timelike coordinate. Concurrent with this curve, for any accelerated quark trajectory there appears on the string worldsheet an event horizon  \cite{dragtime}, which is an indicator in the gravity side of quark energy loss via gluonic radiation.\footnote{The connection between energy loss and the presence of a worldsheet horizon  was noticed initially in \cite{gubserqhat,ctqhat} at finite temperature, and in
  \cite{dragtime} (see also \cite{dominguez,xiao}) for the zero temperature case.}

We also learn from (\ref{mikhwsmetric}) that the constant-$\tau_r$ lines  are null with respect to $g_{ab}$.
 Mikhailov's solution is thus built upon a parametrization of the string worldsheet where the natural notion of proper time $\tau_r$ associated (modulo a rescaling by $R^2/z^2$) with the endpoint is extended to the full worldsheet by following the null geodesics directed toward larger $z$. These null lines at constant $\tau_r$ (or $t_r$) have a clear physical interpretation: they represent the trajectories on the worldsheet along which endpoint information propagates, so in gauge theory language, they encode the contribution to the gluonic field  that the quark at the given retarded time $\tau_r$ generates at any distance $z$ to the source (or equivalently, at any later time $t$).

 It is therefore natural to examine  the question of beaming directly in Mikhailov's parametrization, $\tau=\tau_r$, $\sigma=z$.
 An immediate observation is that (\ref{mikhsolnoncovariant}) stipulates that
 \begin{equation}\label{mikhbeaming}
 z=\sqrt{1-\vec{v}(t_r)^{2}}(t-t_r)~,
 \end{equation}
 which shows that when the quark is ultra-relativistic at some event $(t_r,\vec{x}(t_r))$ along its worldline, the corresponding null line on the string worldsheet automatically remains close to the AdS boundary for a long time.
 This guarantees that the contribution to the chromoelectromagnetic field sourced specifically by the event in question (via the dilatonic or gravitational field set up by the string, as reviewed in Section \ref{mappingsubsec},)  will remain localized even a long distance away from the source. In essence, this specific contribution will be peaked within a narrow cone around the quark velocity, much as in classical electromagnetism, with the opening angle of the cone determined by the slope $1/\gamma$ of the Mikhailov line in question.  We thus see that, \emph{independently of the validity of any approximation scheme, equation (\ref{mikhbeaming}) embodies the beaming of the gluonic (near and radiation) fields emitted by an ultra-relativistic quark}. The overall pattern may or may not be substantially localized, depending on the quark behavior at other times. The effect is illustrated in Fig.~1, for a particular choice of trajectory.

\begin{figure}[htb]
\label{beamingfig}
\centering
\begin{minipage}[t]{7cm}
\includegraphics[width=6.5cm]{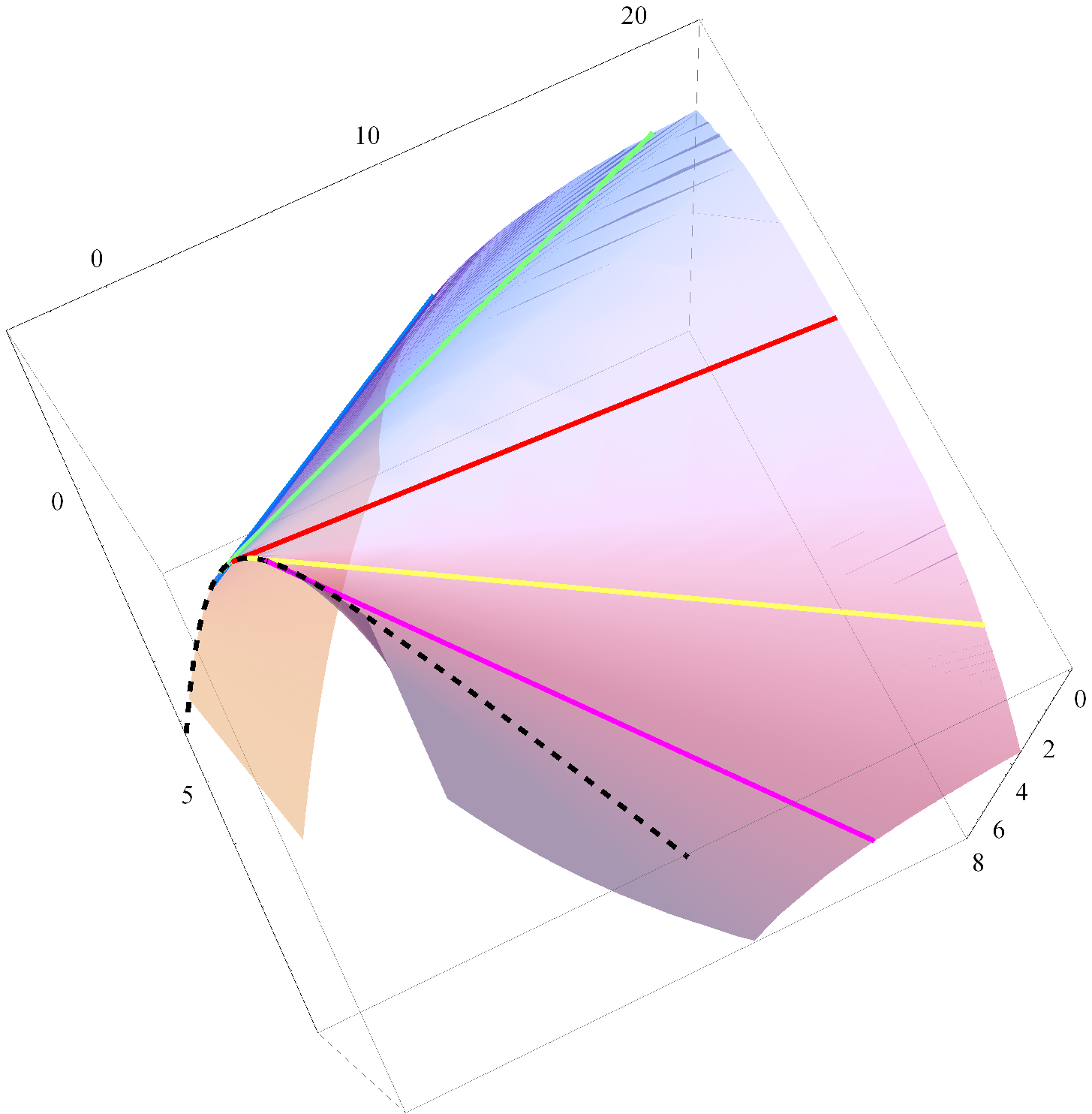}
\end{minipage}
\begin{minipage}[t]{7cm}
\includegraphics[width=6.8cm]{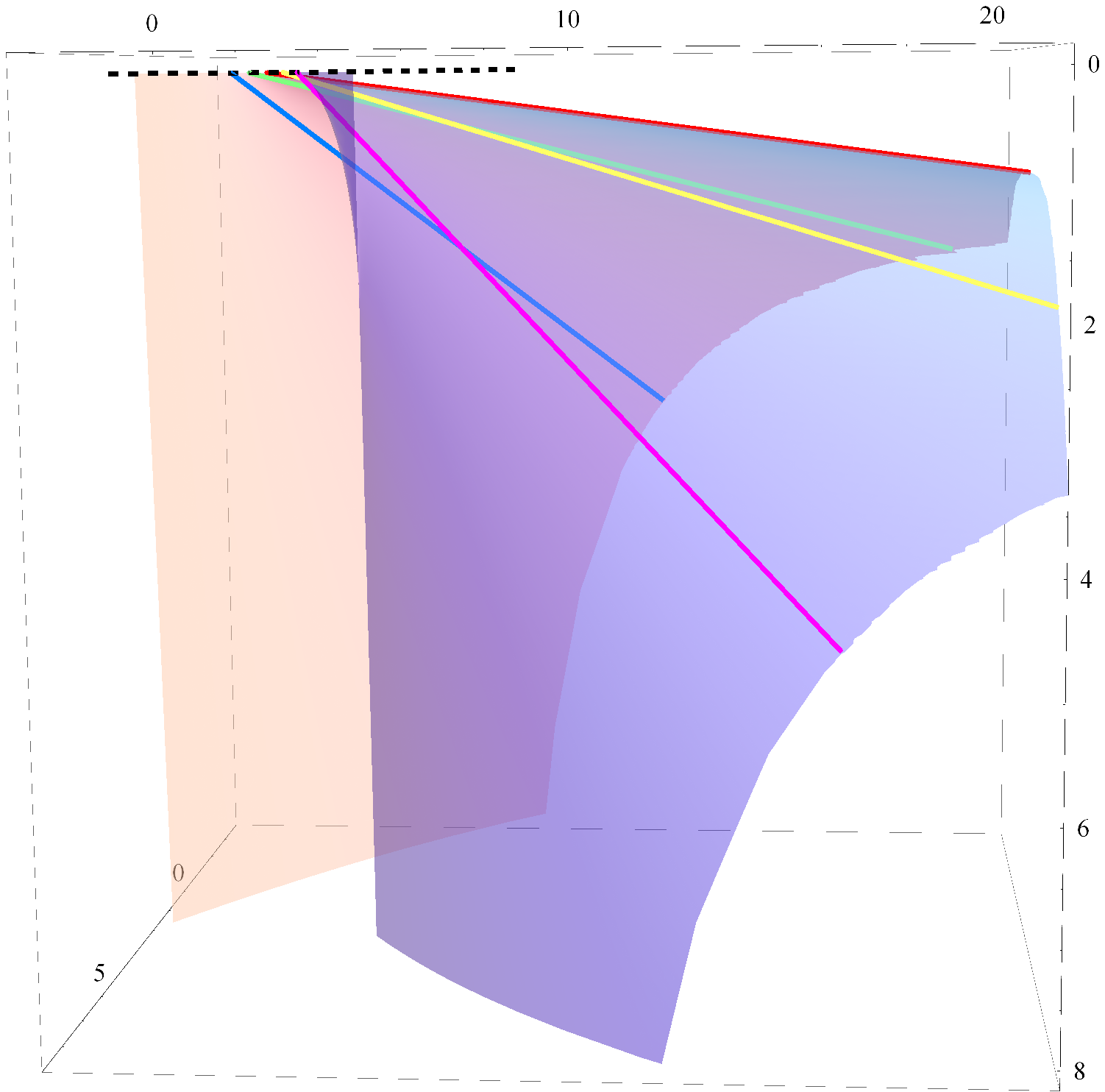}
\end{minipage}
\setlength{\unitlength}{1cm}
\begin{picture}(0,0)
\put(-0.6,4.9){\small $z$}
\put(-0.5,4.6){\vector(0,-1){0.6}}
\put(-5.6,6.2){\small $x_1$}
\put(-5.1,6.3){\vector(1,0){0.6}}
\put(-13.6,5.1){\small $x_1$}
\put(-13.1,5.3){\vector(2,1){0.6}}
\put(-13.9,2){\small $x_2$}
\put(-13.6,1.6){\vector(1,-2){0.3}}
\put(-8.2,1.7){\small $z$}
\put(-8.2,1.5){\vector(-1,-2){0.2}}
\end{picture}
 \caption{To exemplify the beaming effect, the figure on the left (right) depicts a top (side) view of a the string embedding dual to a quark undergoing motion along the path marked with dotted lines, the hyperbola $x_2=\sqrt{c^2+x_1^2}$ (at $x_3=0$), with velocity $v_1(t_r)=v_{\mbox{\scriptsize{max}}}\mbox{sech}(t_r/T)$, for the choice of parameters $c=2$, $T=2$ and
$v_{\mbox{\scriptsize{max}}}=0.999$. The surface shown is swept out by the string as the observation time $t$ (not indicated in the figure) progresses. The straight lines indicate the null trajectories along which endpoint information flows, i.e., the locations on the string that encode the gluonic radiation emitted by the quark at a given retarded time (for $t_r=-\frac{T}{2},-\frac{T}{4},0,\frac{T}{4},\frac{T}{2}$). These lines have slope $1/\gamma$, so as the velocity increases, they remain closer to the AdS boundary (the plane at the top of the figure), implying a larger degree of collimation for the corresponding contribution to the overall gluonic field profile. The line closest to the boundary (red online) emanates from the endpoint at $t_r=0$, when the quark has the maximal velocity $v_{\mbox{\scriptsize{max}}}$.}
\end{figure}

 This property underlies in particular the (partly) unexpected results of \cite{liusynchrotron} for synchrotron radiation. In that case, the uniform circular motion of the quark gives rise to a spiralling string profile, which is a particular instance of (\ref{mikhsolnoncovariant}) but was derived independently in \cite{liusynchrotron}. It was observed in that paper that in the $|\vec{v}|\to 1$ limit, the synchrotron radiation pattern  precisely lines up  with the projection of the spiralling string onto the AdS boundary. We now see that this is explained by (\ref{mikhbeaming}),  through which  the ultra-relativistic limit forces the entire string to approach the boundary. And, most importantly, the connection with the general embedding (\ref{mikhsolnoncovariant}) informs us that the beaming effect is not confined to the case of uniform circular motion, but follows in fact from (\ref{mikhbeaming}) for arbitrary trajectories, at any emission point where the quark is sufficiently relativistic.

 Another simple example is the case of a quark traveling at a constant ultra-relativistic velocity. To avoid potential confusion, we emphasize that the known fact that the corresponding string is always vertical (i.e., extending purely along the radial AdS coordinate $z$) is perfectly compatible with our finding that the associated Mikhailov lines (\ref{mikhbeaming}) are nearly parallel to the AdS boundary (i.e., lie almost at constant $z$, like the $t_r=0$ line in Fig.~1). These are just 2 different ways of slicing the same worldsheet: the first with lines at fixed $t$, the second with lines at fixed $t_r$. In this case, we of course know by Lorentz invariance that the net effect of the beaming will be to localize the gluonic field to the plane that contains the quark and is transverse to its velocity,\footnote{{}Using the results of \cite{iancu2,trfsq}, it is easy to check that the delta function enforcing this localization is multiplied by additional factors which diverge as $|\vec{v}|\to 1$ in the case of $\expec{T_{00}(x)}$ and vanish in the case of  $\expec{\trFsq(x)}$.} yielding a chromoelectromagnetic shock wave.\footnote{As we will elaborate on in Section \ref{shocksubsec}, the proposal of \cite{veronika} makes crucial use of shock waves in the AdS, rather than CFT, fields.}

We should note that the existence of a beaming effect at any ultra-relativistic point of an otherwise general quark worldline is implicit in the results of \cite{iancu2,trfsq}, which respectively worked out the detailed form of $\expec{T_{00}(x)}$ and $\expec{\trFsq(x)}$. The focus of these papers, however, was the surprising fact that the integrated contribution from the entire string can be rewritten as a function of only the string endpoint at a single appropriately retarded time, and the consequent lack of temporal/radial broadening in the gluonic profile. No attempt was made there to examine the origin of the collimation effect, and the unexpected proximity of the string to the AdS boundary embodied in (\ref{mikhbeaming}) was not noticed.

Having understood on the gravity side the physical origin of gluonic beaming for arbitrary quark trajectories, in terms of the unusual proximity of the string with the boundary, in the following sections we will turn our attention to Hubeny's proposal \cite{veronika}. Concretely, we will examine the conditions under which the string can be approximated as a collection of particles that emit bulk fields localized along shock wave fronts.

\section{Transverse String Velocity}\label{transversesec}

A quantity of crucial importance for the beaming calculation of \cite{veronika} is the \emph{transverse} velocity
\begin{equation}\label{vperp}
V^m_{\perp}\equiv\Xd^m-\left(\frac{\Xd\cdot\Xp}{\Xp\cdot\Xp}\right)\Xp^m~,
\end{equation}
 which, for any given choice of time coordinate $\tau$, is a measure of the motion of the individual points of the string that is invariant under reparametrizations of $\sigma$ (including those that are $\tau$-dependent). For use in the next section, we note that the Nambu-Goto Lagrangian (\ref{nambugoto}) can be expressed as\footnote{Incidentally, this relation (which is surely well-known--- see, e.g., \cite{zwiebach}) shows that the force-dependent limiting velocity for a forced finitely-massive quark in a thermal plasma derived in \cite{dragtime} (generalizing previous findings in \cite{argyres1,gubserqhat,ctqhat,mmt,liu4,argyres3}) follows from the simple requirement that the 5-vector $V^m_{\perp}$ of the string endpoint not be spacelike.}
\begin{equation}\label{ngvperp}
\sqrt{\left(\Xd\cdot\Xp\right)^2-\Xd^2\Xp^2}={\sqrt{\Xp^2}}\sqrt{-V^2_{\perp}}~.
\end{equation}

Denoting the spatial components of the transverse velocity as $\overrightarrow{V}_{\perp}\equiv(\vec{V}_{\perp},V^z_{\perp})$, it will be convenient to define
 \begin{equation}\label{vperpspatialnorm}
 \overrightarrow{V}_{\perp}^2 \equiv \vec{V}_{\perp}^2 + (V^z_{\perp})^2~,
 \end{equation}
 omitting the factor of $R^2/z^2$ present in the metric (\ref{metric}).

In static gauge, (\ref{vperp}) says that
\begin{equation}\label{vperpspatial}
\overrightarrow{V}_{\perp}=\frac{1}{1+\vec{X}^{'2}}
\left( (1+\vec{X}^{'2})\dot{\vec{X}}-(\dot{\vec{X}} \cdot\vec{X}^{'})\vec{X}^{'}, -\dot{\vec{X}}\cdot\vec{X}^{'}\right)~,
\end{equation}
and hence
\begin{equation}\label{vperpspatialmag}
\overrightarrow{V}_{\perp}^2 = \dot{\vec{X}}^2 - \frac{(\dot{\vec{X}}\cdot\vec{X}^{'})^2}{1+\vec{X}^{'2}}~.
\end{equation}
Knowing that (\ref{mikhsolnoncovariant}) implies \cite{dragtime}
\begin{eqnarray} \label{Xdot}
\dot{\vec{X}}&\equiv&
\left(\frac{\partial \vec{X}}{\partial t}\right)_{z}
=\vec{v}+\frac{(1-\vec{v}^{\,2})\vec{a}z}{(\vec{v}\cdot\vec{a})z
+(1-\vec{v}^{\,2})^{3/2}}~,
\\
\label{Xprime}
\vec{X}^{'}&\equiv&\left(\frac{\partial\vec{X}}{\partial z}\right)_{t}
=-\frac{\sqrt{1-\vec{v}^{\,2}}\,\vec{a}z}{(\vec{v}\cdot\vec{a})z
+(1-\vec{v}^{\,2})^{3/2}}
\end{eqnarray}
(with the velocity $\vec{v}$ and acceleration $\vec{a}$  evaluated at the appropriate retarded time), we can rewrite (\ref{vperpspatialmag}) as
\begin{equation}\label{vperpsq1}
\overrightarrow{V}^2_{\perp}=1-\frac{(1-\vec{v}^{\,2})^3}{\vec{a}^{\,2}z^2}
\left[\frac{\vec{X}^{'2}}{1+\vec{X}^{'2}}\right]~,
\end{equation}
or, purely in terms of quark data,
\begin{equation} \label{vperpsq}
\overrightarrow{V}^2_{\perp}= 1-\frac{(1-\vec{v}^{\,2})^4}{(1-\vec{v}^{\,2})\vec{a}^{\,2} z^2+\left[(\vec{v}\cdot\vec{a})z+(1-\vec{v}^{\,2})^{3/2}\right]^2}~.
\end{equation}
Notice that, as $\vec{a}\to 0$, (\ref{Xprime}) implies that $\vec{X}^{'}\to 0$ at the rate necessary for (\ref{vperpsq1}) to give the finite result $\overrightarrow{V}^2_{\perp}=\vec{v}^{\,2}$ that is directly visible from (\ref{vperpsq}).

\section{Pulverizing the String}\label{pulverizingsec}
\subsection{$\expec{\trFsq(x)}$ and the dilaton field}\label{fsqsubsec}

 As we have recalled in section \ref{mappingsubsec}, $\expec{\trFsq(x^{\mu})}$ is obtained via (\ref{trfsq}) from the dilaton field sourced by the string. The latter is given by (\ref{dilstring}), or equivalently,
\begin{equation} \label{dilstring2}
\varphi(x)={1\over 4\pi\ls^2}\int d\tau\,d\sigma\,\sqrt{\Xp^2(\tau,\sigma)}\sqrt{-V^2_{\perp}(\tau,\sigma)}
  \,D(x;X(\tau,\sigma))~,
\end{equation}
where we have made use of (\ref{ngvperp}).
Notice that we are purposefully leaving the string parametrization unspecified.
Our interest in this subsection is to determine under what conditions (\ref{dilstring2}) (and therefore (\ref{trfsq})) can be approximated by using point particles in place of the string, as proposed in \cite{veronika}.

 The dilaton field generated by a point particle is deduced by substituting the particle action in place of the Nambu-Goto action in the definition of the source (\ref{stringsource}):
\begin{equation} \label{particlesource}
J(x)\equiv {{R^3}\over (16\pi G^{(5)}_{\mbox{\scriptsize N}})^2}\frac{\delta S_{\mbox{\scriptsize part}}}{\delta \varphi(x)}=\frac{m}{2}
     \int d\tau \sqrt{-V^2(\tau)}\,\delta^{(5)}\left(x-X(\tau)\right)~,
\end{equation}
with $V^m\equiv\Xd^m$ the particle velocity, leading to
\begin{equation} \label{dilpart}
\varphi(x)=\frac{m}{2}\int d\tau\,\sqrt{-V^2(\tau)}
  \,D(x;X(\tau))~.
\end{equation}
Comparing this with (\ref{dilstring2}), it becomes clear that, for the purpose of computing the dilaton field (and consequently, $\expec{\trFsq(x)}$), the string (or a segment of it) can be approximated as a collection of particles moving with velocities $V^m=V^m_{\perp}$ only to the extent that
\begin{equation}\label{xdotcondition}
V^m_{\perp}\simeq \Xd^m
\end{equation}
and
 \begin{equation}\label{xprimecondition}
\Xp^2\simeq\mbox{constant}~.
 \end{equation}
The first condition is forced upon us by the fact that, in the particle expression (\ref{dilpart}), it is certainly true that the source velocity and location are related. If we fail to meet this requirement, then the string expression (\ref{dilstring2}) can at best match onto a version of (\ref{dilpart}) where we treat $V^m$ and $X^m$ as independent variables. {}From a mathematical standpoint, we could still claim in that case to be decomposing the string into pointlike sources, but we would have to bear in mind that the field produced by such sources will \emph{not} coincide with standard particle results.

Condition (\ref{xprimecondition}) is needed if we insist on obtaining particles with a uniform mass density, so that their contributions to the total dilaton field are assigned \emph{equal weight}. (This is in fact what was assumed in \cite{veronika} for the case of the graviton field.) Of course, even when  (\ref{xprimecondition}) is not satisfied, (\ref{ngvperp}) and (\ref{xdotcondition}) would entitle us to assert that the string is equivalent to a collection of particles with masses that depend on time and position. Unfortunately, this description seems unlikely to be of any use, except perhaps in very special cases where such dependence turns out to be simple. To be able to employ standard particle results, we should at least insist on the mass function $\sqrt{\Xp^2}$ being only a function of the particle index $\sigma$ and not of the worldline time $\tau$.

Notice that, in spite of the fact that (\ref{trfsq}) is sensitive only to the near-boundary behavior of the dilaton, our comparison ensures that the original string and its pulverized version generate (approximately) equal dilaton profiles throughout the bulk of AdS. The approach to the boundary, $z\to 0$,  affects only the dilaton propagator $D(x;X(\tau))$, which is common to the string and particle descriptions. In addition, even if we were somehow able to find a collection of particles that yield a dilaton profile matching that of the string only on the AdS boundary, and not in the bulk, it is clear that the corresponding MSYM \emph{states} will be different, and this difference could be detected by computing non-local observables, such as correlators at 2 or more points.

\subsection{$\expec{T_{\mu\nu}(x)}$ and the graviton field}\label{tmunusubsec}

We know from Section \ref{mappingsubsec} that the MSYM energy-momentum tensor is obtained from (\ref{graltmunu}), where the graviton field is sourced by the string energy-momentum tensor (\ref{stringtmn}). The latter can be rewritten in the form
\begin{equation} \label{stringtmn2}
\mathrm{T}^{mn}(x)= \frac{1}{2\pi\ls^2}
     \int  \frac{d^2\sigma}{\sqrt{-G}\sqrt{-V^2_{\perp}}}
  \mathcal{V}^{mn}\,\delta^{(5)}\left(x-X(\tau,\sigma)\right)~,
\end{equation}
where we have used (\ref{ngvperp}) and defined
\begin{equation}\label{vcal}
\mathcal{V}^{mn}\equiv\frac{1}{\sqrt{\Xp^2}}\left[\Xp^2\Xd^m\Xd^n+\Xd^2\Xp^m\Xp^n
-\Xd\cdot\Xp\left(\Xd^m\Xp^n+\Xp^m\Xd^n\right)\right]~.
\end{equation}
{}From the definition (\ref{vperp}) of the transverse velocity, it is easy to show that, in an arbitrary string parametrization,
 \begin{equation}\label{vcal2}
 \mathcal{V}^{mn}=\sqrt{\Xp^2}V^m_{\perp}V^n_{\perp}+\frac{V^2_{\perp}}{\sqrt{\Xp^2}}\Xp^m\Xp^n~.
 \end{equation}
Since the propagator $\mathcal{G}_{mn;\bar{m}\bar{n}}$ is chosen to satisfy the gauge condition $h_{mz}=0$, we know that the graviton field will receive contributions only from the $\mu,\nu$ components of $\mathcal{V}^{mn}$.

On the other hand, the particle energy-momentum tensor is of course
\begin{equation} \label{particletmn}
\mathrm{T}^{mn}(x)\equiv \frac{2}{\sqrt{-G}}\frac{\delta S_{\mbox{\scriptsize part}}}{\delta G_{mn}(x)}=m
     \int  \frac{d\tau\,V^m V^n}{\sqrt{-G}\sqrt{-V^2}}\,\delta^{(5)}\left(x-X(\tau)\right)~.
\end{equation}
Comparing  this against (\ref{stringtmn2}), we learn that the string energy-momentum tensor (and consequently the graviton field it generates, as well as the gauge theory energy-momentum tensor extracted from the latter) can be well approximated by the aggregated effect of a collection of particles moving with velocities $V^m=V^m_{\perp}$ only if conditions (\ref{xdotcondition}) and (\ref{xprimecondition}) are satisfied, together with the requirement that
 \begin{equation}\label{othercondition}
\left| \frac{1}{V^2_{\perp}}V^{\mu}_{\perp}V^{\nu}_{\perp} \right|
\gg
\left| \frac{1}{\Xp^2}\Xp^{\mu}\Xp^{\nu} \right|\quad\forall\quad \mu,\nu~.
 \end{equation}
This demands that the transverse velocity of the string be ultra-relativistic, in the sense that
\begin{equation}\label{relativisticcondition}
 \frac{z^2}{R^2}V^2_{\perp}\simeq 0~.
\end{equation}
 In other words, the spatial norm defined in (\ref{vperpspatialnorm}) must satisfy
$\overrightarrow{V}_{\perp}^2\simeq (V_{\perp}^0)^2$.
The interesting feature here is that the condition of ultra-relativistic motion, which was the focus of \cite{veronika}, has emerged naturally from the requirement that the string be `pulverizable', i.e., replaceable by a collection of particles.

\subsection{Shock waves}\label{shocksubsec}

In the previous 2 subsections, we have learned that, in general, the string can be decomposed into particles in a manageable way only if conditions (\ref{xdotcondition}), (\ref{xprimecondition}) and (\ref{relativisticcondition}) are satisfied. In the remaining sections of the paper, we will examine to what extent these 3 `pulverizability' conditions are in fact fulfilled in different parametrizations of the string worldsheet. Before leaving this section, however, we wish to make another remark.

{}From the beginning, we are working in the linearized approximation (justified at $N_c\gg 1$), so the total field is to be obtained directly as the superposition of the contribution produced by each particle. Condition (\ref{relativisticcondition}) requires that the motion of the resulting particles be ultra-relativistic, and so naturally leads one to expect the bulk (dilatonic or gravitational) field to be well approximated as a superposition of shock waves generated separately by each string bit, as advocated in \cite{veronika}. In the gravitational case, these would be the AdS analogs \cite{hi} of the Aichelburg-Sexl shock waves \cite{as}, obtained by scaling $|\vec{v}|\to 1$ with $\gamma m$ fixed, and would be localized on a hypersurface transverse to the particle velocity. In the dilatonic case, the factors of $V^2$ work out differently (compare (\ref{dilpart}) vs. (\ref{hsol}) and (\ref{particletmn})), so in the strict $|\vec{v}|\to 1$ limit there will exist a finite shock wave only if one holds $m/\gamma$ fixed, but the localization properties are exactly the same. Both cases are in complete analogy with the electromagnetic field of a highly boosted charge in Minkowski space, which is localized to a plane transverse to the particle velocity as a result of Lorentz contraction.\footnote{The same is if course true in the chromoelectromagnetic case, as we noted in Section \ref{originsec}.}  This localization is the essential reason why it is easier to set up the bulk field calculation in terms of shock waves, as was done in \cite{veronika}.

 To examine this feature in a little more detail, recall from the Lienard-Wiechert solution in classical electrodynamics that, in the ultra-relativistic limit, the dominant contribution to the fields will be \emph{beamed} along a direction closely aligned with the particle's velocity. The transverse plane relevant at a given observation time $t$, then, emerges from a sum of beamed contributions emitted at all preceding instants $\bar{t}\le t$ along the particle's worldline, and therefore requires that the velocity remain constant throughout the motion. More precisely, the farther away we wish the fields to be well approximated by a shock wave localized on the plane transverse to the particle's velocity, the longer the period throughout which this velocity must have been essentially constant in order for such localization to arise.
 If this requirement is not met, the locus where the field is peaked will not coincide with the transverse plane.

 For the dilatonic or gravitational fields in the AdS geometry (\ref{metric}), there is an analogous restriction, with the appropriate word replacements. Instead of being required to travel along a straight line at constant speed, the particle will have to move along a nearly null geodesic. For us, this does not entail any change, because, in the Poincar\'e coordinates we are using (where (\ref{metric}) holds), null geodesics in fact are straight lines.
 The curved geometry does imply, however, that the fields generated by the rapidly moving particle will not be localized on a plane, but on a hemisphere transverse to the particle velocity and orthogonal to the AdS boundary, as was shown in \cite{veronika} via a construction involving spatial geodesics.\footnote{More precisely, the shock wave front is exactly a hemisphere orthogonal to $V_{\perp}$ if it is viewed at fixed $t$ and $V_{\perp}$ is computed using $\tau=t$. In other parametrizations, the locus where the fields are peaked would naturally be a distorted version of a hemisphere.} In order for the hemispherical shock wave to have had a chance to form, the bit must have been traveling along a nearly straight line for a sufficiently long time.

 Thus, even if the string satisfies conditions (\ref{xdotcondition}) and (\ref{relativisticcondition}) and can  therefore be decomposed into (automatically ultra-relativistic) particles, and even if it also satisfies (\ref{xprimecondition}) so that these  particles all contribute with equal weight, it is not guaranteed that the bulk fields generated by them are correctly approximated all the way to the AdS boundary by the shock waves considered in \cite{veronika}.  The requirement of shock wave formation is therefore a fourth condition whose validity we need to explore separately from the 3 pulverizability requirements of the previous subsections. We will return to it at the end of the paper.

\section{Pulverization in (Reversed) Mikhailov Gauge}\label{mikhailovsec}

To look for situations where the gluonic field at sufficiently long distances from the quark could become well approximated by the beaming prescription of \cite{veronika}, we would like to find string embeddings where the 3 pulverizability conditions  deduced in the previous section are satisfied with arbitrarily good precision at large $z$, which corresponds to the infrared region of the CFT.

In Appendix A we show that when these conditions are imposed in the Mikhailov gauge $\tau=\tau_r$, $\sigma=z$,  we are in fact forced to restrict attention to a region where $z$ is small (the only exception being the case where the quark has a constant ultra-relativistic velocity). This restriction is related to the fact that, for $z>1/\sqrt{a^2}$, it is $z$ and not $\tau_r$ that plays the role of timelike coordinate. If we want to have a good approximation scheme at large $z$, it is thus natural  to work in the `reversed Mikhailov' gauge $\tau=z$, $\sigma=\tau_r$. In this case the formula (\ref{vperp}) for the transverse velocity gives
\begin{equation}\label{mikhvperp}
V^m_{\perp}=\left(\frac{z^2 a^2 v^{\mu}+za^{\mu}}{z^2 a^2 -1},1\right)~.
\end{equation}
It follows that
\begin{equation}\label{mikhvperpsq}
V^2_{\perp}=\frac{R^2}{z^2(1-z^2 a^2)}~,
\end{equation}
so as expected the string bits follow timelike trajectories only in the region inside the stationary limit curve (\ref{slc}), $z^2>1/a^2$.
The transverse velocity (\ref{mikhvperp}) approaches $\Xd^m=(v^{\mu},1)$  at
\begin{equation}\label{farbeyondslc}
z^2\gg 1/a^2~,
\end{equation}
so the basic particle requirement (\ref{xdotcondition}) is satisfied in this region (which of course exists only for $\tau_r$ such that $a^2>0$). {}From (\ref{mikhvperpsq}) we see that the ultra-relativistic condition (\ref{relativisticcondition})  holds as well, in the same region. These 2 conditions ensure that the portion of the string under examination can be approximated as a collection of particles. In the region of interest $V_{\perp}^m\simeq (v^{\mu},1)$ is independent of the time $z$,
which means that the contributions from the individual string bits can be approximated as (dilatonic or gravitational) shock waves.
On the other hand, for arbitrary quark motion, the equal-weight condition (\ref{xprimecondition}) is \emph{not} satisfied: since $\Xp^2=R^2(z^2 a^2-1)/z^2\simeq R^2 a^2$, our particles have a mass (density) that is independent of the time $z$ but depends on the particle label $\tau_r$. Condition (\ref{xprimecondition}) is fulfilled only to the extent that $a^2$ is approximately constant, and in particular, it hold when $a^2$ is strictly constant, as happens for constant 4-acceleration and for uniform circular motion.

The region where the decomposition into particles is valid, delimited by (\ref{farbeyondslc}),  lies deep beyond the stationary limit curve (\ref{slc}). For quark trajectories where $v^{\mu}$ asymptotes to a constant in the distant past and future, this curve encircles only a limited portion of the worldsheet (part of which lies beyond the associated event horizon) \cite{dragtime}, so the string can be pulverized only in a relatively narrow region. Worse, due to the structure of the embedding (\ref{mikhsolnoncovariant}), this region is generally tilted toward future times, whereas  the swath of the worldsheet that is relevant for computing the gluonic field profile at a given observation point is tilted toward the past, so there cannot be a substantial overlap between the 2. In particular, along this swath the conditions (\ref{xdotcondition}) and (\ref{relativisticcondition}) are \emph{not} satisfied at arbitrarily large values of $z$.

For the restricted class of quark worldlines whose $a^2$ is bounded below by some constant $a_{\mbox{\scriptsize min}}^2$, the string can be decomposed into particles for all $z\gg 1/a_{\mbox{\scriptsize min}}^2$. This includes the cases of uniform circular motion and uniform 4-acceleration, but in the latter case the string is known to reach only up to $z=\sqrt{a^{-2}+t^2}$ before turning back towards the AdS boundary \cite{xiao,brownian,nolineonthehorizon,veronikasemenoff}.

\section{Particle Gauge}\label{particlesec}

As we have just seen, even though Mikhailov's parametrization precisely tracks the flow of quark data along the string worldsheet, and explains through (\ref{mikhbeaming}) the beaming of the gluonic field for ultra-relativistic quark motion, it does not provide a description of the string in terms of particles where all 3 of the pulverizability conditions from Section \ref{pulverizingsec} are usefully satisfied, except in the case of constant ultra-relativistic velocity and in a very restricted class of quark trajectories whose only memorable member is precisely the case studied in \cite{veronika}, uniform circular motion. It is thus natural to wonder if there might exist a different parametrization of the solutions (\ref{mikhsol}) that is better suited to extending the approximation scheme proposed in \cite{veronika} beyond the original example of synchrotron motion.

 Our strategy in this section will be to fix the gauge on the worldsheet using 2 out of the 3 pulverizability conditions (\ref{xdotcondition}), (\ref{xprimecondition}) and (\ref{relativisticcondition}). The first 2 relations are the ones that can be most naturally interpreted as strict equalities, so we will concentrate on them and come back to (\ref{relativisticcondition}) at the end. We stipulate then our choice of worldsheet coordinates $\tau=\btau\,$, $\sigma=\bsigma$ by demanding that:
 \begin{enumerate}
 \item The string bits behave exactly as true particles in the sense of (\ref{xdotcondition}), i.e., $V_{\perp}^m=\Xd^m$, which by (\ref{vperp}) is equivalent to
 \begin{equation}\label{xdotgauge}
 \Xd\cdot\Xp=0~.
 \end{equation}
 We recognize this as one of the 2 conditions imposed in the familiar conformal gauge.
 \item The string bits all have exactly identical mass, in the sense of (\ref{xprimecondition}), so that their contributions to the bulk fields are weighted uniformly. That is,
 \begin{equation}\label{xprimegauge}
 \Xp^2=R^2~,
 \end{equation}
 where we have used our knowledge that $\Xp^{m}$ is always spacelike (because $\Xd^m$ is by construction timelike), and made a convenient choice for  the overall normalization. (Here we deviate from the other conformal gauge condition, $\Xp^2=-\Xd^2$.)
 \end{enumerate}
 These 2 conditions fix the worldsheet coordinates up to $\btau\to f(\btau)$ reparametrizations, and we will call them `particle gauge' (bearing in mind from Section \ref{pulverizingsec} that they suffice to pulverize the string into particles only as far as the dilaton field is concerned).

 Using the explicit form of the string embedding (\ref{mikhsol}), the gauge conditions (\ref{xdotgauge}) and (\ref{xprimegauge}) translate into
 \begin{eqnarray}\label{particlegauge}
(1-z^2 a^2) \tau_r'\dot{\tau_r}+\tau_r'\dot{z}+\dot{\tau_r}z'&=&0~,\nonumber\\
(1-z^2 a^2) \tau_r'^{\,2}+2\tau_r'z'+z^2&=&0~,
 \end{eqnarray}
 or their inverted version\footnote{We let $'$ and $\dot{}$ respectively denote partial derivatives with respect to $\bsigma$ and $\btau$ or $z$ and $\tau_r$, depending on which set of variables is being differentiated.}
\begin{eqnarray}\label{particlegauge2}
(z^2 a^2-1) \btau'\bsigma'+\btau'\dot{\bsigma}+\dot{\btau}\,\bsigma'&=&0~,\nonumber\\
(z^2 a^2-1) \btau'^{\,2}+2\dot{\btau}\,\btau'-z^2(\dot{\btau}\,\bsigma'-\btau'\dot{\bsigma})^2&=&0~.
 \end{eqnarray}
 One can reprocess (\ref{particlegauge}) and (\ref{particlegauge2}) to obtain a relation involving only $\bsigma(\tau_r,z)$,
 \begin{equation}\label{particlegauge3}
 (1-z^2 a^2) \bsigma'^{\,2}-2\dot{\bsigma}\,\bsigma'-\frac{1}{z^2}=0~.
 \end{equation}
 It is easy to convince oneself that these equations always prevent us from choosing $\tau_r'=0$ or $\dot{\tau_r}=0$ or $z'=0$, and generally forbid $\dot{z}=0$ as well, so it is difficult to get explicit expressions for $\btau$ and $\bsigma$ when the quark trajectory is arbitrary.

 As a proof of concept, here we will content ourselves with examining how the remaining pulverizability condition fares for the restricted class of trajectories where $a^2$ is constant, which was already found to be special in the previous section, and includes the 3 textbook cases of interest: constant velocity, uniform circular motion, and  constant 4-acceleration (although the latter suffers from the limitation mentioned at the end of the previous section). For constant $a^2$, starting from (\ref{particlegauge3}) we can derive the explicit solution
 \begin{eqnarray}\label{bars}
 \bsigma(\tau_r,z)&=&b\tau_r-\ln\left(\frac{cz}{1+\sqrt{1+(b^2-a^2)z^2}}\right)\\
 {}&{}&\quad\;\;\,
 +\frac{b}{2a}\ln\left(\frac{a^2(az+1)^2}{a^2+b^2+(b^2-a^2)a^2 z^2 + 2 ab\sqrt{1+(b^2-a^2)z^2}}\right)~,
 \nonumber\\
 \btau(\tau_r,z)&=&d\tau_r
 +\frac{d}{2a}\ln\left(\frac{a^2(az+1)^2}{a^2+b^2+(b^2-a^2)a^2 z^2 + 2 ab\sqrt{1+(b^2-a^2)z^2}}\right)~,
 \nonumber
 \end{eqnarray}
with $b,c,d$ integration constants.
Notice that this solution is well-defined both outside and inside the stationary limit curve (\ref{slc}),
which in this case is static and therefore coincides with the worldsheet event horizon. For simplicity, we choose $b=c=d=a$, for which the relations reduce to
\begin{eqnarray}\label{bars2}
 \bsigma(\tau_r,z)&=&a\tau_r
 +\ln\left(\frac{az+1}{2az}\right)~,
\\
 \btau(\tau_r,z)&=&a\tau_r
  +\ln\left(\frac{az+1}{2}\right)~.
 \nonumber
\end{eqnarray}

Let us now turn our attention to the remaining condition. {}From (\ref{bars2}) we can work out
\begin{equation}\label{vperpsqparticlegauge2}
V_{\perp}^2=-\frac{R^2}{a^2 z^2}~,
\end{equation}
which unfortunately does not satisfy the ultra-relativistic condition (\ref{relativisticcondition}).\footnote{It is not possible to  avoid this negative conclusion by adjusting the integration constants $b,c,d$ in (\ref{bars}) in a different manner.} That is, the particles into which we have pulverized the string \emph{do not} become arbitrarily relativistic as we move to larger values of $z$. {}From our discussion in Section \ref{tmunusubsec}, we know this implies that, in this gauge, the gravitational field sourced by the string (and consequently the stress-energy tensor sourced by the quark) cannot be obtained as a sum of individual particle contributions. It is also easy to see that the $V_{\perp}$ is not nearly constant,
so, while the dilaton field sourced by the string (and the dual MSYM Lagrangian density sourced by the quark) can be approximated in terms of equally-weighted particles, this approximation cannot invoke shock waves.

Having failed to satisfy the 3 desired conditions even in the $a^2=\mbox{const.}$ case, which was about the only one where we had  succeeded in the previous section, we harbor little hope that the particle gauge might prove useful in practice to extend the shock wave prescription of \cite{veronika} to more general quark trajectories.

\section{(Semi-)Static Gauge} \label{staticsec}

\subsection{Static gauge} \label{staticsubsec}

As explained in the Introduction, the shock wave prescription was successfully put to the test in \cite{veronika}, for the special case of uniform circular motion, in a calculation that directly employed the natural static gauge, $\tau=t$, $\sigma=z$. It is therefore interesting to examine to what extent the 3 pulverizability conditions of Section \ref{pulverizingsec} hold in this gauge, for arbitrary trajectories, in the regime of large $z$ or, equivalently, large distances away from the source.

 The transverse velocity in static gauge was computed already in Section \ref{transversesec}. Looking at (\ref{vperpspatial}), the first thing we notice is that condition (\ref{xdotcondition}), which embodies the essential connection between the position and velocity of the putative particles,  is satisfied only if $(\dot{\vec{X}} \cdot\vec{X}^{'})^2\ll \dot{\vec{X}}^2 [1 +(\vec{X}^{'})^2 ]$. Given the form of (\ref{Xdot}) and (\ref{Xprime}), it is easy to see that this requirement is automatically satisfied when the quark itself is ultra-relativistic, except in the case of worldlines that asymptote at early times to uniform circular motion. {}From (\ref{vperpsq}) we see that in this case the string bits are also ultra-relativistic, in the sense of (\ref{relativisticcondition}), for all $z$. These 2 conditions are enough to ensure that the string can be decomposed into particles, but for a generic trajectory (\ref{xprimecondition}) is not satisfied, so these particles cannot be assigned equal weight. Worse, the mass density that would have to be assigned to the collection of particles is in general time-dependent (even at large $z$), which means that in determining the resulting bulk fields one cannot make use of standard particle results. An exception is the case of asymptotic uniform circular motion, whose very special features lead to the fulfillment of (\ref{xprimecondition}) for all $z\gg 1/\gamma^2 |\vec{a}|$. Another special case is the one of constant velocity, where the mass density is simply $\mu=R/2\pi\ls^2 z$, and therefore easily manageable.\footnote{As mentioned in Section \ref{originsec}, it is easy to see from the exact results of \cite{iancu2,trfsq} that the gluonic field collapses to the plane transverse to $\vec{v}$, as is of course expected by Lorentz contraction.}

 For non-relativistic quark motion, (\ref{xdotcondition}) is fulfilled only for a very limited class of quark trajectories, including the case of constant velocity, but not the case of uniform circular motion. The other 2 conditions would impose additional restrictions on the trajectory. In Appendix B we show that, curiously, conditions (\ref{xprimecondition}) and (\ref{relativisticcondition}) by themselves are only satisfied simultaneously by quark worldlines that asymptote at early times to uniform circular motion, precisely the case considered in \cite{veronika}.

We conclude then that, whether or not the quark is ultra-relativistic, if we wish to somehow justify the calculation that was carried in \cite{veronika} for the synchrotron case, we will have to deviate from this parametrization of the worldsheet.

\subsection{Semi-static gauge} \label{semistaticsubsec}
As we have just seen, for general quark trajectories, including the case of uniform circular motion, the static gauge violates the essential particle condition (\ref{xdotcondition}). How can we explain, then, the successful treatment in \cite{veronika} of the case where the quark undergoes uniform circular motion, which was carried out precisely in static gauge? The answer lies in the fact that the essential ingredient for the calculation of \cite{veronika} was the transverse string velocity (\ref{vperpspatial}), which is by construction invariant under arbitrary reparametrizations of the spatial worldsheet coordinate. We are thus led to keep $\tau=t$ but seek a new $\sigma=\zeta(t,z)$ such that ({\ref{xdotcondition}}) is satisfied, without modifying (\ref{vperpspatial}).

The partial derivatives with respect to the new coordinates $(t,\zeta)$ will evidently be related to their static gauge counterparts via
\begin{eqnarray}
\label{xdotsigma}
 \left(\frac{\partial{X}}{\partial t}\right)_{\zeta}&=&\dot{X} + X^{'} \left(\frac{\partial z}{\partial t}\right)_{\zeta}~,\\
 \label{xprimesigma}
 \left(\frac{\partial{X}}{\partial \zeta}\right)_{t}&=&X^{'} \left(\frac{\partial z}{\partial \zeta }\right)_{t}~,
\end{eqnarray}
where for uniform circular motion
\begin{eqnarray}
\dot{{X}}&=&(1,\vec{v}+\gamma\vec{a}z,0)~,
\\
{X}^{'}&=&(0,-\gamma^2\vec{a}z,1)~.
\end{eqnarray}
Condition ({\ref{xdotcondition}}) is thus seen to hold exactly if
\begin{eqnarray}
 \dot{z}\equiv \left(\frac{\partial z}{\partial t}\right)_{\zeta}=-\frac{\dot{X}\cdot{X}^{'}}{{X}^{'}\cdot {X}^{'}}~.
\end{eqnarray}
For uniform circular motion this says that
\begin{eqnarray}
 \dot{z}=\frac{\gamma^3{\vec{a}}^{\,2}z^2}{1+\gamma^4{\vec{a}}^{\,2}z^2}~,
\end{eqnarray}
and integrating, we obtain
\begin{eqnarray}
\label{sigmastaticgauge}
 t-f(\zeta)= \gamma z- \frac{1}{\gamma^3{\vec{a}}^{\,2}z}~.
\end{eqnarray}
where $f(\zeta)$ is an arbitrary function of the new coordinate $\zeta$, which is defined by this last equation.

We thus see that it is not at all difficult to find a parametrization where (\ref{xdotcondition}) is satisfied. We will use the remaining freedom to demand that the equal mass requirement (\ref{xprimecondition}) also be fulfilled.
It follows from (\ref{sigmastaticgauge}) that
\begin{eqnarray}\label{zprimesemistatic}
z^{'}\equiv \left(\frac{\partial z}{\partial  \zeta}\right)_{t}=\frac{f^{'}(\zeta)\gamma^3{\vec{a}}^{\,2} z^2}{1+\gamma^4{\vec{a}}^{\,2} z^2}~,
\end{eqnarray}
so we can  see that
\begin{eqnarray}
\sqrt{\left(\frac{\partial X}{\partial \zeta}\right)_{t}^{2}}=\sqrt{\frac{R^2}{z^2}({1+\gamma^4{\vec{a}}^{\,2} z^2})}\,z^{'}
\end{eqnarray}
will satisfy (\ref{xprimecondition}) only if $z^{'}\simeq \mbox{constant}$ at large $z$, or more specifically, for
\begin{equation}\label{ultraregion}
z\gg 1/\gamma^2|\vec{a}|~,
 \end{equation}
 which is the same region that had been encountered in (\ref{farbeyondslc}) and in the second paragraph of the previous subsection.
 {}From (\ref{zprimesemistatic}), this entails that $f^{'}(\zeta)$ is also constant. This requirement eliminates the main arbitrariness of our new coordinate system, and  $\zeta$  is now determined up to 2 integration constants, which we set to convenient values:
\begin{eqnarray}
\label{sigmastaticgauge2}
 \zeta=  t-\gamma z+ \frac{1}{\gamma^3{\vec{a}}^{\,2}z}~.
\end{eqnarray}
Inverting this, we find
\begin{eqnarray} \label{sigmastaticgauge3}
 z=\frac{t-\zeta}{2\gamma}+\sqrt{\left(\frac{t-\zeta}{2\gamma}\right)^2+\frac{1}{\gamma^4{\vec{a}}^{\,2}}}~.
\end{eqnarray}
 We refer to this new parametrization of the worldsheet as `semi-static' gauge.

\begin{figure}[h]
\centering
\begin{minipage}[t]{7.5cm}
\includegraphics[width=7.4cm]{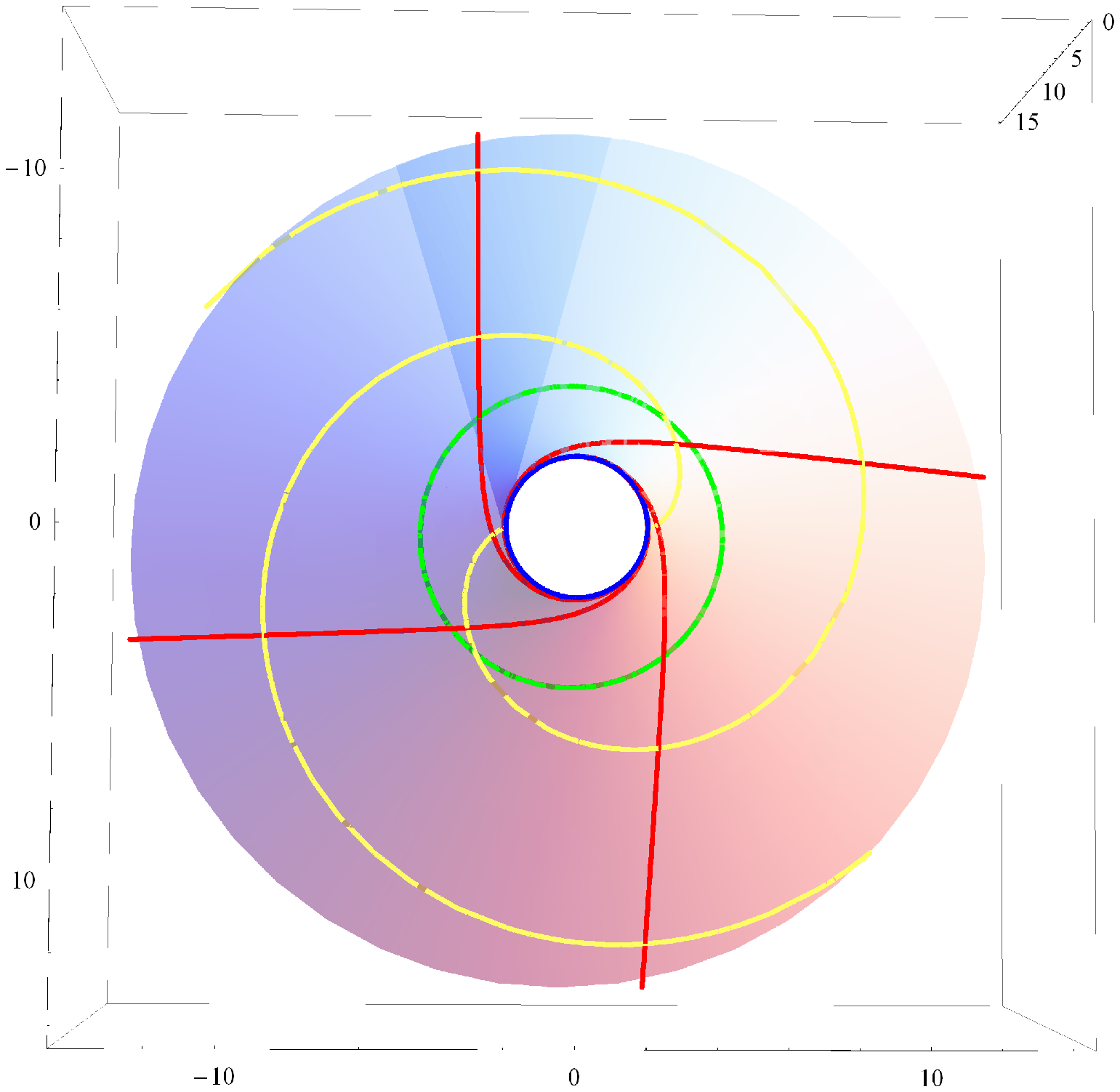}
\end{minipage}
\begin{minipage}[t]{7.5cm}
\includegraphics[width=7.0cm]{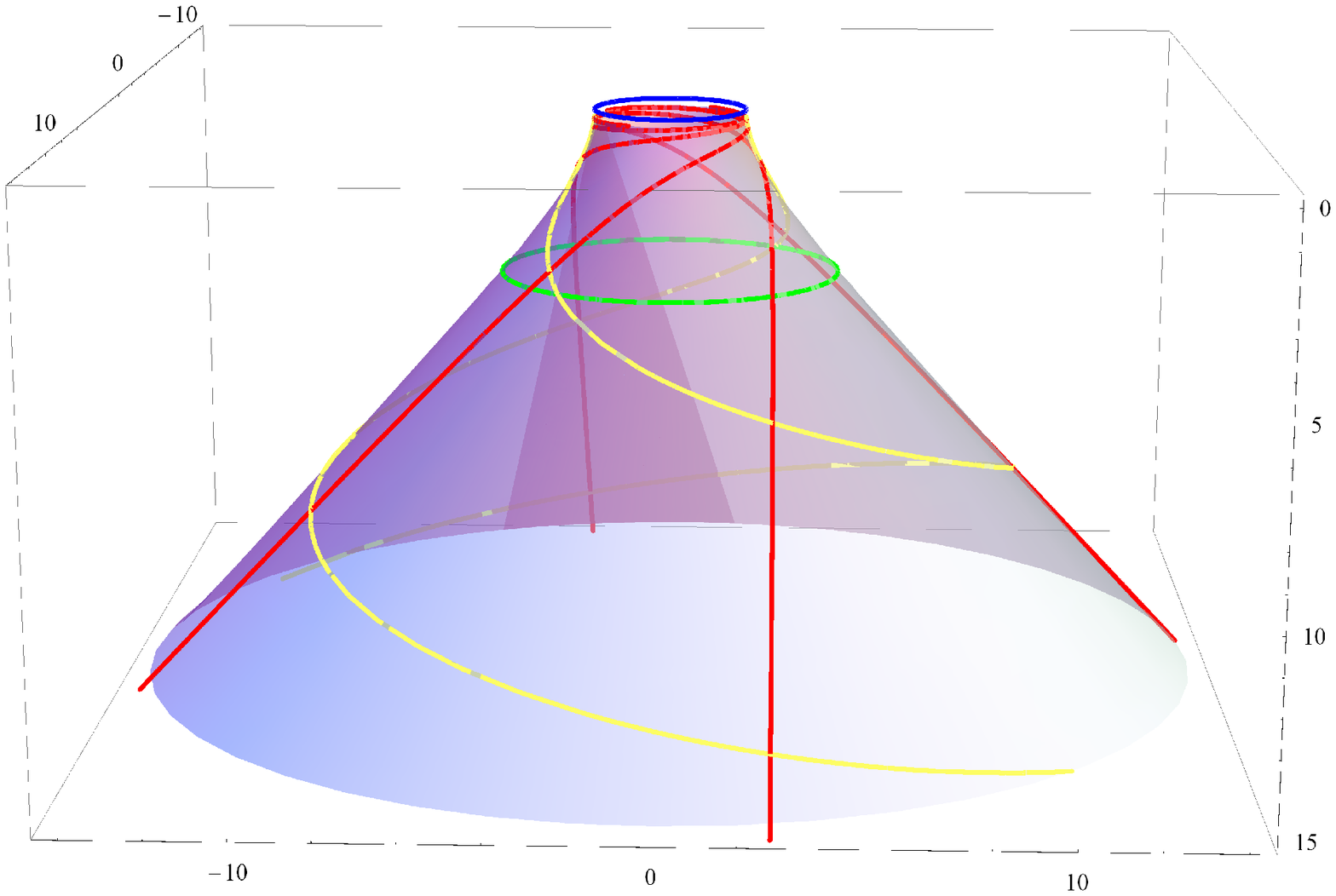}
\end{minipage}
\setlength{\unitlength}{1cm}
\begin{picture}(0,0)
\put(-1.3,6.4){\small $z$}
\put(-1.2,6.3){\vector(-1,-1){0.4}}
\put(-3.5,0.7){\small $x_1$}
\put(-3.0,0.8){\vector(1,0){0.6}}
\put(-6.7,3.8){\small $x_2$}
\put(-6.6,3.4){\vector(0,-1){0.6}}
\put(6.8,4.5){\small $z$}
\put(6.9,4.2){\vector(0,-1){0.6}}
\put(3.2,1.0){\small $x_1$}
\put(3.7,1.1){\vector(1,0){0.6}}
\put(0.5,5.6){\small $x_2$}
\put(0.6,5.4){\vector(-2,-1){0.6}}
\end{picture}
\label{figura3} \caption{Side and top views of the conical worldsheet swept out by the string dual to a quark in uniform circular motion (with speed $v=0.68$ and radius $R=1$). In the figure on the left (right), the trajectory of the quark/string endpoint is indicated by the innermost (top) circle (blue online). The circle near the middle of the figures (green online) indicates the location of the worldsheet horizon. The spiral with uniform pitch (yellow online) represents the profile of the string $\vec{X}(t,z)$ at a given time, which rotates rigidly as time progresses. Four additional curves are shown (red online) that first repeatedly circle the origin forming a very tightly wound spiral (with pitch decreasing to zero as $z\to 0$), and then open up into nearly straight lines. These portray $\vec{X}(t,\zeta)$ at fixed values of $\zeta$, and so correspond to the trajectories of 4 of the point particles into which the string can be decomposed, moving with velocity ${V}_{\perp}$. See main text for discussion.
}
\end{figure}

To get a sense of this new slicing of the string embedding, fix an initial time $t=t_i$, and consider the point on the string  at some given initial radial position $z=z_i>0$. Through (\ref{sigmastaticgauge2}), this defines a corresponding initial $\zeta=\zeta_i$. If we follow this point (i.e., hold $\sigma$ fixed) as time evolves, then in static gauge ($\sigma=z$) it just circles around at the same radial depth, while in semi-static gauge ($\sigma=\zeta$) it traces a spiral that goes ever deeper into AdS, according to (\ref{sigmastaticgauge3}), with ever-increasing pitch. See Figure~\ref{figura3}. Varying $\zeta_i$ (equivalently, $t_i$) we sweep out the entire worldsheet\footnote{Except for the point at $z=0$, for which the transformation (\ref{sigmastaticgauge2}) is ill-defined.}  using a collection of spiraling worldlines (related to each other through rotation) which all pass through the same plane $z=z_i$  with different spatial velocities $\overrightarrow{V}_{\perp}$ given by (\ref{vperpspatial}) at the corresponding retarded time $t_{r,i}=\zeta_i - 1/
 \gamma^3{\vec{a}}^{\,2}z_i$. As expected, the new parametrization is periodic with respect to $\zeta$.

By construction, in semi-static gauge our pulverizability conditions (\ref{xdotcondition}) and (\ref{xprimecondition}) are satisfied in the region (\ref{ultraregion}), i.e., for the portion of the string far beyond the stationary limit curve (\ref{slc}), which in the case under consideration, coincides with the worldsheet event horizon. These 2 conditions guarantee that the resulting assemblage of particles gives a good approximation of the dilaton field sourced by the string. {}From (\ref{vperpsq}) it is clear that, independently of the value of $|\vec{v}|$,  the ultra-relativistic condition (\ref{relativisticcondition}) is also satisfied in the region (\ref{ultraregion}) (as was shown already in \cite{veronika}), which means that the graviton field is also well approximated with the same particles.

We conclude then that, in the case of uniform circular motion, the use of semi-static gauge allows all 3 of our pulverizability conditions to be satisfied at large $z$. This is consistent  with the success of the calculation performed in \cite{veronika}. At the same time, the analysis in this and the previous sections makes it clear that the fulfillment of the 3 conditions (\ref{xdotcondition}), (\ref{xprimecondition}) and (\ref{relativisticcondition}) hinges on various special features of circular motion, and unfortunately does not extend to arbitrary quark trajectories.

\subsection{Shock wave formation condition}
\label{shockformationsubsec}

Now that we know that the spiralling string dual to a quark in uniform circular motion can be decomposed into ultra-relativistic particles of equal weight, let us finally return to the shock wave formation condition identified in Section \ref{shocksubsec}.
As we explained there, we need to check whether the velocity of each particle asymptotes to a constant, which is necessary for the hemispherical
 shock waves of \cite{veronika} to correctly approximate the particle's contribution to the dilatonic and gravitational fields. The analysis is easy to carry out for the present  case of uniform circular motion (and would apply equally if the quark only approached this type of motion asymptotically at early times). The transverse velocity (\ref{vperpspatial}) takes the form
\begin{eqnarray}
\overrightarrow{V}_{\perp}=\left(\vec{v}+\frac{\gamma \vec{a}z}{1+\gamma^4{\vec{a}}^{\,2}z^2},\frac{\gamma^3{\vec{a}}^{\,2}z^2}{1+\gamma^4{\vec{a}}^{\,2}z^2}\right)~,
\end{eqnarray}
which, in the region (\ref{ultraregion}), simplifies to
\begin{eqnarray}\label{vperpultraregion}
\overrightarrow{V}_{\perp}\simeq \left(\vec{v}+\frac{\vec{a}}{\gamma^3{\vec{a}}^{\,2}z},\frac{1}{\gamma}
-\frac{1}{\gamma^5{\vec{a}}^{\,2}z^2}\right)~.
\end{eqnarray}
We see here that at large $z$ the 3-velocity of each particle along the directions parallel to the AdS boundary is governed mainly by the velocity of the quark at the retarded time $t_r$, while the $z$ component of its velocity tends to a constant value. Since $\vec{v}$ rotates, (\ref{vperpultraregion}) might seem at first sight incompatible with the requirement of having a nearly constant velocity.
 Fortunately, the incompatibility is only apparent, because the variation of the transverse velocity with respect to the time $t$ is appropriately small, in spite of the fact that it is not small with respect to the retarded time.

To see this, we can vary equation (\ref{mikhsolnoncovariant}) and use the $z$-component of the transverse velocity to deduce that
\begin{eqnarray}
 dz&=&V^z_{\perp}dt\simeq \frac{dt}{\gamma}~,\nonumber \\
dt_r&=&\frac{dt}{1+\gamma^4{\vec{a}}^{\,2}z^2}\simeq \frac{dt}{\gamma^4{\vec{a}}^{\,2}z^2}~.
\end{eqnarray}
Then, the variation of the transverse velocity to leading order in the region (\ref{ultraregion}) is
\begin{eqnarray}
d \overrightarrow{V}_{\perp} \simeq \left(-\frac{ \vec{v}}{\gamma^7{\vec{a}}^{\,2}\vec{v}^{\,2}z^3},\frac{2}{\gamma^6{\vec{a}}^{\,2}z^3}\right)dt~,
\end{eqnarray}
which is indeed small in the region of interest.

It is useful to develop this point a bit further. We can express the transverse velocity in terms of the quark's variables evaluated at the retarded time $t_r^\infty$ associated with $z\to \infty$ via a series expansion,
\begin{eqnarray}
\vec{v}(t_r)&\simeq& \vec{v}_\infty+\vec{a}_{\infty}(t_r-t_r^{\infty})-\frac{1}{2}\omega^2 \vec{v}_\infty(t_r-t_r^{\infty})^2~,\\
\vec{a}(t_r)&\simeq &\vec{a}_{\infty}-\omega^2 \vec{v}_{\infty}(t_r-t_r^{\infty})~, \nonumber
\end{eqnarray}
where the subindex $\infty$ implies that those variables are evaluated at $t_r^\infty$. From (\ref{sigmastaticgauge2})  we can see that
\begin{eqnarray}
 t_r-t_r^{\infty}=-\frac{1}{\gamma^3 \vec{a}^{\,2} z}~,
\end{eqnarray}
 is small compared to the period $T$ in the region of interest, and therefore allows the series expansion. This expressions also tells us that, contrary to what a quick glance at (\ref{vperpultraregion}) might suggest, each particle actually reaches $z\to\infty$ before completing even one turn of the spiral (a fact that is also evident in Fig.~2).
Using these approximations, we can express the transverse velocity (\ref{vperpultraregion}) as
\begin{eqnarray}
\overrightarrow{V}_{\perp}\simeq \left(\vec{v}_{\infty}+\frac{{\vec{v}}_{\infty}}{2\gamma^6{\vec{a}}^{\,2}z^2\vec{v}^{\,2}_{\infty}},\frac{1}{\gamma}
-\frac{1}{\gamma^5{\vec{a}}^{\,2}z^2}\right)~.
\end{eqnarray}
We see here very explicitly that the variation of the transverse velocity from its asymptotic value is small in the stated region, as is necessary for the associated hemispherical shock wave to emerge.

 The only remaining concern is that, as described in Section \ref{shocksubsec}, the scheme of \cite{veronika} assumes that the string bit in question has a transverse velocity that is well approximated by its asymptotic (null) value for a time that is sufficiently long to build up the shock wave hemisphere down to its equator at the AdS boundary. By considering the delay in propagation, it is easy to see that this actually amounts to artificially extending the bit's trajectory along a straight line all the way back to $z=0$, instead of including the real (tightly wound spiral) motion shown in Fig.~2. When the speed of the quark approaches 1, the portion of the string bit's trajectory that deviates from this extrapolation is pushed towards $z=0$, which explains why the calculation performed in \cite{veronika} for the synchrotron radiation pattern worked so well for $\gamma\gg 1$.  But for mildly relativistic or non-relativistic quark speeds, the hemispherical shock waves are not reliable all the way to the AdS boundary. It is conceivable that in the special case of uniform circular motion (and for the graviton field needed for the observable $\expec{T_{00}(x)}$ scrutinized in \cite{veronika}), the error induced by the incorrect extension of the string bits' trajectories is small, but we do not see any obvious reason why this should be the case. At any rate, the analysis of the present paper has taught us that, unfortunately, the shock wave prescription of \cite{veronika} is not valid for much more general quark trajectories.

\section*{Acknowledgements}
We are grateful to Mariano Chernicoff and Veronika Hubeny for useful discussions and comments on a draft of this paper.
The present work was partially supported by Mexico's National Council of Science and Technology (CONACyT) grant 104649, as well as DGAPA-UNAM grant IN110312.
The research of B.L. was generously supported by funds from the Abdus Salam International Centre for Theoretical Physics.

\appendix
\section*{Appendices}

\section{Pulverization in Mikhailov gauge}

As seen in (\ref{mikhwsmetric}), $\Xp^2=0$ in the Mikhailov gauge $\tau=\tau_r$, $\sigma=z$, so the transverse velocity (\ref{vperp}) is not well-defined. We must then revisit the analysis of Section \ref{pulverizingsec} to derive the analogs of the pulverizability conditions (\ref{xdotcondition}), (\ref{xprimecondition}) and (\ref{relativisticcondition}). Using (\ref{mikhwsmetric}), we see that
 \begin{equation}\label{mikhailovnambugoto}
 \sqrt{\left(\Xd\cdot\Xp\right)^2-\Xd^2\Xp^2}=|\Xd\cdot\Xp|=\frac{R^2}{z^2}~,
 \end{equation}
so the dilaton field (\ref{dilstring2}) is given by
\begin{equation} \label{dilstringmikhailov}
\varphi(x)={1\over 4\pi\ls^2}\int dz \frac{R^2}{z^2} d\tau_r\,D(x;X(\tau_r,z))~.
\end{equation}
There are a number of ways in which we could try to mimic this with a collection of particles at different values of $z$. The most obvious is to take all particles to move with the same velocity $V^m=(v^{\mu},0)$, where $v^{\mu}(\tau_r)$ denotes the quark's 4-velocity at the given retarded time. In this case $\sqrt{-V^2}=R/z$, so (\ref{dilpart}) yields
\begin{equation} \label{dilpartmikhailov}
\varphi(x)=\frac{1}{2}\int dz\frac{\mu R}{z}d\tau_r\,D(x;X(\tau_r,z))~,
\end{equation}
with $\mu$ the mass density of our continuum of particles. The missing factor of $R/z$ in (\ref{dilpartmikhailov}) as compared to (\ref{dilstringmikhailov}) must be supplied by taking
\begin{equation}\label{massdensity}
\mu=\frac{R}{2\pi\ls^2 z}~.
\end{equation}
The contributions of the different particles will then be modulated by this factor, so we do not satisfy the equal-weight requirement that in other gauges led to our condition (\ref{xprimecondition}). Still, we see that in Mikhailov gauge the scaling of the mass is simple enough to be manageable, and in particular, it does not depend on time, or on the form of the quark trajectory.

A more serious problem is that $\Xd^m=(v^{\mu}+za^{\mu},0)$ differs from the velocity $V^m$ we selected to match onto the Nambu-Goto action, so that in general we do not satisfy the analog of (\ref{xdotcondition}), and the string bits cannot properly be regarded as particles, because their positions and velocities are independent data. This problem is only avoided if
\begin{equation}\label{smallzacondition}
z a^{\mu}\ll v^{\mu}~.
\end{equation}

So far we have discussed the conditions relevant to the calculation of the dilaton field. For the graviton field, we know from Section \ref{tmunusubsec} that we need the $\mu,\nu$ components of the string stress tensor (\ref{stringtmn}) to be well approximated by the corresponding components of the particle stress-energy tensor (\ref{particletmn}), i.e.,
\begin{equation}\label{newvcal}
\frac{\Xp^2\Xd^{\mu}\Xd^{\nu}+\Xd^2\Xp^{\mu}\Xp^{\nu}
-\Xd\cdot\Xp\left(\Xd^{\mu}\Xp^{\nu}+\Xp^{\mu}\Xd^{\nu}\right)}{2\pi\ls^2\sqrt{\left(\Xd\cdot\Xp\right)^2-\Xd^2\Xp^2}}
\simeq  \frac{\mu V^{\mu} V^{\nu}}{\sqrt{-V^2}}~.
\end{equation}
Using (\ref{mikhwsmetric}), the left-hand side reduces to
$$
\frac{1}{2\pi\ls^2}\left(v^{\mu}v^{\nu}(1+z^2 a^2) + z a^{\mu} v^{\nu} + z v^{\mu} a^{\nu}\right)~,
$$
so we see that (\ref{newvcal}) is satisfied automatically if the mass density scales as in (\ref{massdensity}) and (\ref{smallzacondition}) holds. In terms of the 3-velocity and 3-acceleration of the quark, the latter condition reads
\begin{equation}\label{smallzacondition2}
(t-t_r)(\gamma^2 \vec{v}\cdot\vec{a},\vec{a}+\gamma^2\vec{v}\cdot\vec{a}\vec{v})\ll(1,\vec{v})~,
\end{equation}
where we have made use of (\ref{mikhbeaming}). In general, this would only be satisfied away from the long-distance (large $z$, and therefore large $t-t_r$) regime that is of interest to us, and working at high quark velocity (large $\gamma$) only makes the situation worse.

The only obvious exception is the case of constant velocity, where $\Xd^m=(v^{\mu},0)=V^m$ identically, and bulk shock waves become relevant if the quark is moving ultra-relativistically, thereby providing an example where Hubeny's mechanism strengthens the beaming effect identified in Section \ref{originsec}.

There are of course other ways in which one might try to pulverize the string in Mikhailov gauge. E.g., one can satisfy (the analog of) (\ref{xdotcondition}) by directly choosing the particle velocity as $V^m=(v^{\mu}+za^{\mu},0)$. In that case $\sqrt{-V^2}=R\sqrt{1/z^2-a^2}$~, which is real only in the region outside the stationary limit curve (\ref{slc}). This again prevents us from exploring the long-distance regime.

\section{Static gauge analysis of 2 pulverizability conditions}

In the static gauge $\tau=t$, $\sigma=z$, the requirement that the slope (\ref{Xprime}) satisfies the equal-weight condition (\ref{xprimecondition}) translates into $|\vec{X}^{'}|\simeq C z$, i.e.,
\begin{equation} \label{xprimecondition2}
\sqrt{1-{\vec{v}}^{\,2}}|\vec{a}|\simeq C[(\vec{v}\cdot\vec{a})z+(1-\vec{v}^{\,2})^{3/2}],
\end{equation}
where $C\neq 0$ is a constant, and equality is understood to hold up to subleading terms at large $z$. Notice in particular that $|\vec{a}|$ can be arbitrarily small, but must be different from zero. Given this behavior for $\vec{X}^{'}$, the requirement that the transverse velocity (\ref{vperpsq1}) be ultrarelativistic amounts simply to
\begin{equation}\label{relativisticcondition2}
\vec{a}^{\,2}z^2\gg(1-\vec{v}^{\,2})^3~.
\end{equation}

If we descend into the bulk by increasing $z$ along the line on the worldsheet at constant $t_r$, then (\ref{relativisticcondition2}) is automatically satisfied as long as $\vec{a}\neq 0$, but clearly (\ref{xprimecondition2}) cannot hold except in the special case $\vec{v}\cdot\vec{a}=0$, $|\vec{a}|=C(1-\vec{v}^{\,2})$. Since $\vec{v}\cdot\vec{a}=(1/2)d\vec{v}^{\,2}/dt_r$, this means that, at the given $t_r$, the quark is executing uniform circular motion on a circle of radius $\rho=\vec{v}^{\,2}/C(1-\vec{v}^{\,2})$.

A more relevant question is whether the string is pulverizable when we consider its shape at fixed $t$. In this case, as we vary $z$ the retarded time must evolve according to (\ref{mikhsolnoncovariant}),
\begin{equation}\label{zmikh}
z=\sqrt{1-\vec{v}^{\,2}}(t-t_r)~.
\end{equation}
The IR region $z\to\infty$ thus corresponds to $t_r\to-\infty$. Using (\ref{zmikh}), our conditions (\ref{xprimecondition2}) and (\ref{relativisticcondition2}) respectively read
\begin{equation}\label{xprimecondition3}
|\vec{a}|\simeq C[\vec{v}\cdot\vec{a}(t-t_r)+(1-\vec{v}^{\,2})]
\end{equation}
and
\begin{equation}\label{relativisticcondition3}
|\vec{a}|(t-t_r)\gg (1-{\vec{v}}^{\,2})~.
\end{equation}
Since $|\vec{a}|$ is bounded, the first of these relations informs us that we must have $\vec{v}\cdot\vec{a}\to 0$ at least as fast as $1/(-t_r)$. This in turn means that $|\vec{v}|$ is approaching a constant value, and from (\ref{relativisticcondition3}) one can argue that this constant cannot be zero.

Employing (\ref{xprimecondition3}) in (\ref{relativisticcondition3}), as well as the identities
$|\vec{a}|=|{d\vec{v}}/{dt_r}|$, and $\vec{v}\cdot\vec{a}=|\vec{v}|{d|\vec{v}|}/{dt_r}$, we get
\begin{equation}
\Big|\frac{d\vec{v}}{dt_r}\Big|(t-t_r)
\gg
\frac{1}{C}\Big|\frac{d\vec{v}}{dt_r}\Big|-|\vec{v}|\frac{d|\vec{v}|}{dt_r}(t-t_r)~.
\end{equation}
Clearly for large $|t_r|$ the first term in the right-hand side is negligible compared to the left-hand side, so our condition boils down to
\begin{equation} \label{beamredcondition}
\Big|\frac{d\vec{v}}{dt_r}\Big|
\gg
|\vec{v}|\Big|\frac{d|\vec{v}|}{dt_r}\Big|~,
\end{equation}
where we have eliminated the common positive factor $(t-t_r)$. We have noted above that $|\vec{v}|$ cannot be approaching zero, so the essential requirement is really just
\begin{equation}\label{beamsimply}
\Big|\frac{d\vec{v}}{dt_r}\Big|
\gg
\Big|\frac{d|\vec{v}|}{dt_r}\Big|~.
\end{equation}

In order to interpret this last condition, it is convenient to separate $\vec{v}=|\vec{v}|\hat{u}$, where $\hat{u}$ is a unitary vector in the direction of the velocity. Computing the norm of the acceleration vector we obtain
\begin{equation}
\Big|\frac{d\vec{v}}{dt_r}\Big|
=
\Bigg|\frac{d|\vec{v}|}{dt_r}\hat{u}+|\vec{v}|\frac{d\hat{u}}{dt_r}\Bigg|
\leq
\Big|\frac{d|\vec{v}|}{dt_r}\Big|+|\vec{v}|\Big|\frac{d\hat{u}}{dt_r}\Big|~,
\end{equation}
where we have used the triangle inequality. It is therefore always true that
\begin{equation} \label{trian}
\Big|\frac{d|\vec{v}|}{dt_r}\Big|
\geq
\Big|\frac{d\vec{v}}{dt_r}\Big|-|\vec{v}| \Big|\frac{d\hat{u}}{dt_r}\Big|~,
\end{equation}
so the only way that (\ref{beamsimply}) can hold is if
\begin{equation}
\Big|\frac{d\vec{v}}{dt_r}\Big|
\simeq
|\vec{v}| \Big|\frac{d\hat{u}}{dt_r}\Big|~.
\end{equation}
In other words, asymptotically, the quark must be undergoing uniform circular motion.

The preceding analysis is already very informative, but if our ultimate interest is to compute the gluonic profile set up by the quark, we must remember that the value of $\expec{\trFsq}$ or $\expec{T_{\mu\nu}}$ at a given observation point is determined by combining the dilatonic or gravitational field sourced by the individual points along the string at \emph{different times}. Concretely, the emission point on the string $(t,\vec{X}(t,z),z)$ and the observation point $(t_{\scriptsize\mbox{obs}},\vec{x}_{\scriptsize\mbox{obs}},0)$ are connected through a null geodesic \cite{dkk,trfsq},
\begin{equation}\label{null}
-(t_{\scriptsize\mbox{obs}}-t)^2+(\vec{x}_{\scriptsize\mbox{obs}}-\vec{X}(t,z))^2+z^2=0~,
\end{equation}
which via our knowledge of the explicit form (\ref{mikhsolnoncovariant}) of the string embedding translates into the relation
\begin{equation}\label{zobs}
z=\frac{\sqrt{1-\vec{v}^{\,2}}\left[(t_{\scriptsize\mbox{obs}}-t_r)^2
-(\vec{x}_{\scriptsize\mbox{obs}}-\vec{x})^2\right]}
{2\left[t_{\scriptsize\mbox{obs}}-t_r-(\vec{x}_{\scriptsize\mbox{obs}}-\vec{x})\cdot\vec{v}\,\right]}~.
\end{equation}
On the right hand side, the quark position $\vec{x}$ and velocity $\vec{v}$ are as usual understood to be evaluated at the retarded time $t_r$. As we go to the IR region $t_r\to-\infty$ holding  $t_{\scriptsize\mbox{obs}}$ and $\vec{x}_{\scriptsize\mbox{obs}}$ fixed, $\vec{x}$ may or may not be bounded, and (\ref{zobs}) simplifies to
\begin{equation}\label{zobs2}
z\simeq \frac{\sqrt{1-\vec{v}^{\,2}}(t_r^2 -\vec{x}^{\,2})}
{2(-t_r+\vec{x}\cdot\vec{v}\,)}~.
\end{equation}
For motions where the quark does not approach the speed of light at early times, the denominator cannot vanish, and $\vec{x}$ can grow at most as fast as $\vec{v}\,t_r$. The right-hand side of (\ref{zobs2}) is then of order $(-t_r)$, just like in (\ref{zmikh}). If the quark asymptotically approaches the speed of light, the factor of $t_r^2 -\vec{x}^{\,2}$ in the numerator might be replaced by the subleading expression in (\ref{zobs}), which is of order $-t_r$. But in this case the denominator is likewise of reduced order $(-t_r)^0$, so the overall behavior of $z$ is still the same as in (\ref{zmikh}). Either way, the pulverizability analysis runs exactly in parallel with that in the previous subsection.

We thus conclude that, in static gauge, conditions (\ref{xprimecondition}) and (\ref{relativisticcondition}) are only satisfied to extent that the quark motion at early times asymptotes to uniform circular motion, which was precisely the case studied in \cite{veronika}.


\begin{thebibliography}{99}

 \bibitem{malda}
  J.~M.~Maldacena,
  ``The large $N$ limit of superconformal field theories and
  supergravity,''
  Adv.\ Theor.\ Math.\ Phys.\  {\bf 2}, 231 (1998)
  [Int.\ J.\ Theor.\ Phys.\  {\bf 38}, 1113 (1999)]
  [arXiv:hep-th/9711200].

\bibitem{gkpw}
  S.~S.~Gubser, I.~R.~Klebanov and A.~M.~Polyakov,
  ``Gauge theory correlators from non-critical string theory,''
  Phys.\ Lett.\ B {\bf 428}, 105 (1998)
  [arXiv:hep-th/9802109];\\
  E.~Witten,
  ``Anti-de Sitter space and holography,''
  Adv.\ Theor.\ Math.\ Phys.\  {\bf 2}, 253 (1998)
  [arXiv:hep-th/9802150].

  \bibitem{magoo}
  O.~Aharony, S.~S.~Gubser, J.~M.~Maldacena, H.~Ooguri and Y.~Oz,
   ``Large $N$ field theories, string theory and gravity,''
  Phys.\ Rept.\  {\bf 323}, 183 (2000)
  [arXiv:hep-th/9905111].

\bibitem{cg}
  C.~G.~Callan and A.~G\"uijosa,
  ``Undulating strings and gauge theory waves,''
  Nucl.\ Phys.\ B {\bf 565}, 157 (2000)
  [arXiv:hep-th/9906153].

  \bibitem{mo}
  K.~Maeda and T.~Okamura,
  ``Radiation from an accelerated quark in AdS/CFT,''
  arXiv:0712.4120 [hep-th].

\bibitem{branching}
  Y.~Hatta, E.~Iancu, A.~H.~Mueller,
  ``Jet evolution in the N=4 SYM plasma at strong coupling,''
  JHEP {\bf 0805 } (2008)  037.
  [arXiv:0803.2481 [hep-th]];\\
  Y.~Hatta, T.~Matsuo,
  ``Jet fragmentation and gauge/string duality,''
  Phys.\ Lett.\  {\bf B670 } (2008)  150-153.
  [arXiv:0804.4733 [hep-th]];\\
  E.~Avsar, E.~Iancu, L.~McLerran, D.~N.~Triantafyllopoulos,
  ``Shockwaves and deep inelastic scattering within the gauge/gravity duality,''
  JHEP {\bf 0911 } (2009)  105.
  [arXiv:0907.4604 [hep-th]];\\
  B.~Muller,
  ``Parton Energy Loss in Strongly Coupled AdS/CFT,''
  Nucl.\ Phys.\  {\bf A855 } (2011)  74-82.
  [arXiv:1010.4258 [hep-ph]];\\
  E.~Iancu,
  ``Parton branching and medium-induced radiation in a strongly coupled plasma,''
  Nucl.\ Phys.\  {\bf A855 } (2011)  331-334.
  [arXiv:1012.3527 [hep-ph]].

\bibitem{liusynchrotron}
  C.~Athanasiou, P.~M.~Chesler, H.~Liu, D.~Nickel and K.~Rajagopal,
  ``Synchrotron radiation in strongly coupled conformal field theories,''
  Phys.\ Rev.\  D {\bf 81} (2010) 126001
  [arXiv:1001.3880 [hep-th]].

  \bibitem{veronika}
  V.~E.~Hubeny,
  ``Relativistic Beaming in AdS/CFT,''
  arXiv:1011.1270 [hep-th];\\
  ``Holographic dual of collimated radiation,''
  New J.\ Phys.\  {\bf 13}, 035006 (2011)
  [arXiv:1012.3561 [hep-th]].

\bibitem{hkkky}
  C.~P.~Herzog, A.~Karch, P.~Kovtun, C.~Kozcaz and L.~G.~Yaffe,
  ``Energy loss of a heavy quark moving through $\cN = 4$ supersymmetric
  Yang-Mills plasma,''
  JHEP {\bf 0607} (2006) 013
  [arXiv:hep-th/0605158].

\bibitem{gubser}
  S.~S.~Gubser,
  ``Drag force in AdS/CFT,''
  Phys.\ Rev.\  D {\bf 74} (2006) 126005
  [arXiv:hep-th/0605182].

\bibitem{ct}
  J.~Casalderrey-Solana and D.~Teaney,
  ``Heavy quark diffusion in strongly coupled $\cN = 4$ Yang Mills,''
  Phys.\ Rev.\  D {\bf 74} (2006) 085012
  [arXiv:hep-ph/0605199].

  \bibitem{liuqhat}
  H.~Liu, K.~Rajagopal and U.~A.~Wiedemann,
  ``Calculating the jet quenching parameter from AdS/CFT,''
  Phys.\ Rev.\ Lett.\  {\bf 97} (2006) 182301
  [arXiv:hep-ph/0605178].

\bibitem{clmrw}
  J.~Casalderrey-Solana, H.~Liu, D.~Mateos, K.~Rajagopal and U.~A.~Wiedemann,
  ``Gauge/String Duality, Hot QCD and Heavy Ion Collisions,''
  arXiv:1101.0618 [hep-th];\\
  V.~E.~Hubeny and M.~Rangamani,
  ``A holographic view on physics out of equilibrium,''
  Adv.\ High Energy Phys.\  {\bf 2010} (2010) 297916
  [arXiv:1006.3675 [hep-th]].

\bibitem{gluonicprofile}
  J.~J.~Friess, S.~S.~Gubser, G.~Michalogiorgakis and S.~S.~Pufu,
  ``The stress tensor of a quark moving through $\cN = 4$ thermal plasma,''
  Phys.\ Rev.\  D {\bf 75} (2007) 106003
  [arXiv:hep-th/0607022];\\
  S.~S.~Gubser and S.~S.~Pufu,
  ``Master field treatment of metric perturbations sourced by the trailing
  string,''
  Nucl.\ Phys.\  B {\bf 790} (2008) 42
  [arXiv:hep-th/0703090];\\
  A.~Yarom,
  ``On the energy deposited by a quark moving in an N=4 SYM plasma,''
  Phys.\ Rev.\  D {\bf 75} (2007) 105023
  [arXiv:hep-th/0703095];\\
  S.~S.~Gubser, S.~S.~Pufu and A.~Yarom,
  ``Energy disturbances due to a moving quark from gauge-string duality,''
  JHEP {\bf 0709} (2007) 108
  [arXiv:0706.0213 [hep-th]];\\
  P.~M.~Chesler and L.~G.~Yaffe,
  ``The wake of a quark moving through a strongly-coupled $\mathcal N=4$
  supersymmetric Yang-Mills plasma,''
  Phys.\ Rev.\ Lett.\  {\bf 99} (2007) 152001
  [arXiv:0706.0368 [hep-th]];\\
  S.~S.~Gubser, S.~S.~Pufu and A.~Yarom,
  ``Sonic booms and diffusion wakes generated by a heavy quark in thermal
  AdS/CFT,''
  Phys.\ Rev.\ Lett.\  {\bf 100} (2008) 012301
  [arXiv:0706.4307 [hep-th]];\\
  S.~S.~Gubser and A.~Yarom,
  ``Universality of the diffusion wake in the gauge-string duality,''
  arXiv:0709.1089 [hep-th];\\
  S.~S.~Gubser, S.~S.~Pufu and A.~Yarom,
  ``Shock waves from heavy-quark mesons in AdS/CFT,''
  arXiv:0711.1415 [hep-th];\\
  P.~M.~Chesler and L.~G.~Yaffe,
  ``The stress-energy tensor of a quark moving through a strongly-coupled N=4
  supersymmetric Yang-Mills plasma: comparing hydrodynamics and AdS/CFT,''
  Phys.\ Rev.\  D {\bf 78} (2008) 045013
  [arXiv:0712.0050 [hep-th]];\\
  J.~Noronha, G.~Torrieri and M.~Gyulassy,
  ``Near Zone Navier-Stokes Analysis of Heavy Quark Jet Quenching in an
  $\mathcal{N} =4$ SYM Plasma,''
  arXiv:0712.1053 [hep-ph].



\bibitem{rajagopalshining}
  P.~M.~Chesler, Y.~-Y.~Ho and K.~Rajagopal,
  ``Shining a Gluon Beam Through Quark-Gluon Plasma,''
  arXiv:1111.1691 [hep-th].

\bibitem{mikhailov}
  A.~Mikhailov,
  ``Nonlinear waves in AdS/CFT correspondence,''
  arXiv:hep-th/0305196.

\bibitem{dragtime}
  M.~Chernicoff and A.~G\"uijosa,
  ``Acceleration, Energy Loss and Screening in Strongly-Coupled Gauge
  Theories,''
  JHEP {\bf 0806}, 005 (2008)
  [arXiv:0803.3070 [hep-th]].

  \bibitem{uvir}
  L.~Susskind and E.~Witten,
  ``The Holographic Bound In Anti-De Sitter Space,''
  arXiv:hep-th/9805114;\\
  A.~W.~Peet and J.~Polchinski,
  ``UV/IR relations in AdS dynamics,''
  Phys.\ Rev.\ D {\bf 59} (1999) 065011
  [arXiv:hep-th/9809022].

\bibitem{lorentzdirac}
  M.~Chernicoff, J.~A.~Garc\'\i a and A.~G\"uijosa,
  ``Generalized Lorentz-Dirac Equation for a Strongly-Coupled Gauge Theory,''
 Phys.\ Rev.\ Lett.\  {\bf 102} (2009) 241601
  [arXiv:0903.2047 [hep-th]].

  \bibitem{damping}
  M.~Chernicoff, J.~A.~Garc\'\i a and A.~G\"uijosa,
  ``A Tail of a Quark in $\cN=4$ SYM,''
  JHEP {\bf 0909} (2009) 080
  [arXiv:0906.1592 [hep-th]].
  
  \bibitem{jphysg}
  M.~Chernicoff, J.~A.~Garc\'\i a, A.~G\"uijosa and J.~F.~Pedraza,
  ``Holographic Lessons for Quark Dynamics,''
  J.\ Phys.\ G G {\bf 39} (2012) 054002
  [arXiv:1111.0872 [hep-th]].

  \bibitem{dkk}
  U.~H.~Danielsson, E.~Keski-Vakkuri and M.~Kruczenski,
  ``Vacua, propagators, and holographic probes in AdS/CFT,''
  JHEP {\bf 9901}, 002 (1999)
  [arXiv:hep-th/9812007].

  \bibitem{otherdilaton}
  J.~J.~Friess, S.~S.~Gubser and G.~Michalogiorgakis,
  ``Dissipation from a heavy quark moving through N=4 super-Yang-Mills plasma,''
  JHEP {\bf 0609} (2006) 072
  [hep-th/0605292];\\
  Y.~h.~Gao, W.~S.~Xu and D.~F.~Zeng,
  ``Wake of color fields in charged $\cN = 4$ SYM plasmas,''
  arXiv:hep-th/0606266;\\
  A.~Yarom,
  ``The High momentum behavior of a quark wake,''
  Phys.\ Rev.\ D {\bf 75} (2007) 125010
  [hep-th/0702164].

  \bibitem{trfsq}
  M.~Chernicoff, A.~G\"uijosa and J.~F.~Pedraza,
  ``The Gluonic Field of a Heavy Quark in Conformal Field Theories at Strong Coupling,''
   JHEP {\bf 1110}, 041 (2011)
  [arXiv:1106.4059 [hep-th]].

\bibitem{igor}
  I.~R.~Klebanov,
  ``World volume approach to absorption by nondilatonic branes,''
  Nucl.\ Phys.\  B {\bf 496} (1997) 231
  [arXiv:hep-th/9702076].

\bibitem{iwm}
  I.~R.~Klebanov, W.~Taylor and M.~Van Raamsdonk,
  ``Absorption of dilaton partial waves by D3-branes,''
  Nucl.\ Phys.\  B {\bf 560}, 207 (1999)
  [arXiv:hep-th/9905174].

  \bibitem{shuryakdipole}
  S.~Lin and E.~Shuryak,
  ``Stress tensor of static dipoles in strongly coupled N = 4 gauge theory,''
  Phys.\ Rev.\ D {\bf 76}, 085014 (2007)
  [arXiv:0707.3135 [hep-th]].

\bibitem{iancu1}
  Y.~Hatta, E.~Iancu, A.~H.~Mueller, D.~N.~Triantafyllopoulos,
  ``Aspects of the UV/IR correspondence : energy broadening and string fluctuations,''
  JHEP {\bf 1102 } (2011)  065.
  [arXiv:1011.3763 [hep-th]].

\bibitem{iancu2}
  Y.~Hatta, E.~Iancu, A.~H.~Mueller, D.~N.~Triantafyllopoulos,
  ``Radiation by a heavy quark in N=4 SYM at strong coupling,''
  [arXiv:1102.0232 [hep-th]].

  \bibitem{bk}
  V.~Balasubramanian and P.~Kraus,
  ``A stress tensor for anti-de Sitter gravity,''
  Commun.\ Math.\ Phys.\  {\bf 208} (1999) 413
  [arXiv:hep-th/9902121].

  \bibitem{dhss}
  S.~de Haro, S.~N.~Solodukhin and K.~Skenderis,
  ``Holographic reconstruction of spacetime and renormalization in the  AdS/CFT
  correspondence,''
  Commun.\ Math.\ Phys.\  {\bf 217} (2001) 595
  [arXiv:hep-th/0002230];\\
  K.~Skenderis,
  ``Asymptotically anti-de Sitter spacetimes and their stress energy  tensor,''
  Int.\ J.\ Mod.\ Phys.\  A {\bf 16}, 740 (2001)
  [arXiv:hep-th/0010138].

  \bibitem{dhoker}
  E.~D'Hoker, D.~Z.~Freedman, S.~D.~Mathur, A.~Matusis and L.~Rastelli,
  ``Graviton and gauge boson propagators in AdS(d+1),''
  Nucl.\ Phys.\ B {\bf 562} (1999) 330
  [hep-th/9902042].

  \bibitem{gubserqhat}
  S.~S.~Gubser,
  ``Momentum fluctuations of heavy quarks in the gauge-string duality,''
  Nucl.\ Phys.\  B {\bf 790} (2008) 175
  [arXiv:hep-th/0612143].

\bibitem{ctqhat}
   J.~Casalderrey-Solana and D.~Teaney,
  ``Transverse momentum broadening of a fast quark in a N = 4 Yang Mills
  plasma,''
  JHEP {\bf 0704} (2007) 039
  [arXiv:hep-th/0701123].

  \bibitem{dominguez}
  F.~Dominguez, C.~Marquet, A.~H.~Mueller, B.~Wu and B.~W.~Xiao,
  ``Comparing energy loss and $p_{\perp}$-broadening in perturbative QCD with
  strong coupling $\mathcal{N}=4$ SYM theory,''
  Nucl.\ Phys.\  A {\bf 811} (2008) 197
  [arXiv:0803.3234 [nucl-th]].

\bibitem{xiao}
  B.~W.~Xiao,
  ``On the exact solution of the accelerating string in $AdS_5$ space,''
  Phys.\ Lett.\  B {\bf 665} (2008) 173
  [arXiv:0804.1343 [hep-th]].

\bibitem{zwiebach}
  B.~Zwiebach,
  ``A First Course in String Theory,''
  Cambridge, UK: Univ. Pr. (2009), 673 pp.

  \bibitem{argyres1}
  P.~C.~Argyres, M.~Edalati and J.~F.~V\'azquez-Poritz,
  ``No-drag string configurations for steadily moving quark-antiquark pairs in
  a thermal bath,''
  JHEP {\bf 0701} (2007) 105
  [arXiv:hep-th/0608118].

 \bibitem{mmt}
  D.~Mateos, R.~C.~Myers and R.~M.~Thomson,
  ``Thermodynamics of the brane,''
  JHEP\ {\bf 0705} (2007) 067
  [hep-th/0701132].

    \bibitem{liu4}
  Q.~J.~Ejaz, T.~Faulkner, H.~Liu, K.~Rajagopal and U.~A.~Wiedemann,
  ``A limiting velocity for quarkonium propagation in a strongly coupled plasma
  via AdS/CFT,''
  JHEP\ {\bf 0804}, 089  (2008)
  [arXiv:0712.0590 [hep-th]].

   \bibitem{argyres3}
  P.~C.~Argyres, M.~Edalati and J.~F.~Vazquez-Poritz,
  ``Lightlike Wilson loops from AdS/CFT,''
  JHEP\ {\bf 0803} (2008) 071
  [arXiv:0801.4594 [hep-th]].

  \bibitem{hi}
   M.~Hotta and M.~Tanaka,
  ``Shock wave geometry with nonvanishing cosmological constant,''
  Class.\ Quant.\ Grav.\  {\bf 10} (1993) 307;\\
   G.~T.~Horowitz and N.~Itzhaki,
  ``Black holes, shock waves, and causality in the AdS / CFT correspondence,''
  JHEP {\bf 9902} (1999) 010
  [hep-th/9901012].

  \bibitem{as}
  P.~C.~Aichelburg and R.~U.~Sexl,
  ``On the Gravitational field of a massless particle,''
  Gen.\ Rel.\ Grav.\  {\bf 2}, 303 (1971).


\bibitem{brownian}
  E.~C\'aceres, M.~Chernicoff, A.~G\"uijosa and J.~F.~Pedraza,
  ``Quantum Fluctuations and the Unruh Effect in Strongly-Coupled Conformal
  Field Theories,''
  JHEP {\bf 1006} (2010) 078
  [arXiv:1003.5332 [hep-th]].

  \bibitem{nolineonthehorizon}
   J.~A.~Garc\'{\i}a, A.~G\"uijosa and E.~J.~Pulido,
  in preparation.

    \bibitem{veronikasemenoff}
   V.~E.~Hubeny and G.~W.~Semenoff,
  in preparation.

\bibitem{martinfsq}
  J.~L.~Hovdebo, M.~Kruczenski, D.~Mateos, R.~C.~Myers and D.~J.~Winters,
  ``Holographic mesons: Adding flavor to the AdS/CFT duality,''
  Int.\ J.\ Mod.\ Phys.\  A {\bf 20} (2005) 3428.




 \end{thebibliography}
\end{document}